\documentclass[12pt]{article}
\usepackage{epsfig, color}
\usepackage{a4}
\usepackage{amsmath}
\usepackage{array}
\usepackage{amsfonts}

\newcommand{\ga}{\alpha} 
\newcommand{\gb}{\beta} 
\renewcommand{\gg}{\gamma} 
\newcommand{\gd}{\delta} 
\renewcommand{\ge}{\epsilon}

\newcommand{\gvf}{\varphi} 
\newcommand{\gc}{\chi} 
\newcommand{\gx}{\xi} 
\newcommand{\gm}{\mu}  
\newcommand{\gn}{\nu} 
 
\newcommand{\gk}{\kappa} 
\newcommand{\gl}{\lambda} 
\newcommand{\gr}{\rho} 
\newcommand{\gth}{\theta} 
\newcommand{\gs}{\sigma} 
 
\newcommand{\go}{\omega}

\newcommand{\gps}{\psi} 
 
\newcommand{\gch}{\chi} 
\newcommand{\gG}{\Gamma} 
\newcommand{\gD}{\Delta} 
\newcommand{\gF}{\Phi}

\newcommand{\gL}{\Lambda} 
\newcommand{\gS}{\Sigma} 
 
\newcommand{\gO}{\Omega} 
 
\newcommand{\gPs}{\Psi} 
 
\newcommand{\cK}{{\cal K}} 
\newcommand{\cL}{{\cal L}} 
\newcommand{\cM}{{\cal M}}

\newcommand{\uI}{{\underline I}} 
\newcommand{\uJ}{{\underline J}}

\newcommand{\tk}{{\tilde k}}

\newcommand{\tz}{{\tilde z}} 
\newcommand{\tA}{{\tilde A}}

\newcommand{\tM}{{\tilde M}} 
\newcommand{\tN}{{\tilde N}}

\newcommand{\tZ}{{\tilde Z}}  
\newcommand{\uga}{{\underline\alpha}} 
\newcommand{\ugb}{{\underline\beta}} 
 
\newcommand{\ugd}{{\underline\delta}}

\newcommand{\ugs}{{\underline\sigma}}

\newcommand{\ba}{{\bar a}}

\newcommand{\bh}{{\bar h}}

\newcommand{\bk}{{\bar k}}

\newcommand{\bz}{{\bar z}} 
\newcommand{\bA}{{\bar A}}

\newcommand{\bD}{{\bar D}} 
 
\newcommand{\bF}{{\bar F}}

\newcommand{\bM}{{\bar M}} 
\newcommand{\bN}{{\bar N}}

\newcommand{\bR}{{\bar R}}

\newcommand{\bW}{{\bar W}}

\newcommand{\bZ}{{\bar Z}} 
\newcommand{\bge}{{\bar\epsilon}}

\newcommand{\bgvf}{{\bar\varphi}} 
\newcommand{\bgc}{{\bar\chi}}

\newcommand{\bgl}{{\bar\lambda}}

\newcommand{\bgo}{{\bar\omega}}

\newcommand{\bgps}{{\bar\psi}} 
 
\newcommand{\bgch}{{\bar\chi}} 
\newcommand{\bgF}{{\bar\Phi}}

\newcommand{\bgL}{{\bar\Lambda}} 
\newcommand{\bgS}{{\bar\Sigma}} 
 
\newcommand{\bgO}{{\bar\Omega}} 
 
\newcommand{\bgPs}{{\bar\Psi}} 
 
\newcommand{\bcM}{{\bar{\cal M}}}

\newcommand{\Tr}{\mbox{Tr}} 
\newcommand{\tr}{\mbox{tr}} 
 
\newcommand{\Id}{\scalebox{1}[.95]{1}\hspace{-3.5pt}{\scalebox{1}[1.1]{1}}}
\newcommand{\slashed}{\hspace{-1.1ex}/} 
\newcommand{\Slashed}{\hspace{-1.5ex}/\hspace{.6ex}} 
\newcommand{\lra}{\longrightarrow}

\newcommand{\der}{\partial} 

\newcommand{\Der}{D} 
\newcommand{\sDer}{\Der\Slashed} 
\newcommand{\sder}{\der\slashed}

\newcommand{\nit}{\noindent} 
\newcommand{\nl}{\newline} 
\newcommand{\np}{\newpage} 
\newcommand{\dsp}{\displaystyle}

\newcommand{\ct}{\cite} 
\newcommand{\bit}{\bibitem}

\newcommand{\undr}[1]{{\underline{#1}}} 
\newcommand{\ovr}[1]{{\overline{#1}}} 
\newcommand{\vs}[1]{\vspace{#1 ex}}  

\newcommand{\labl}[1]{\label{#1}} 

\newcommand{\ESO}{$E_6/SO(10)\times U(1)$} 
\newcommand{\Kh}{K\"{a}hler}  
\newcommand{\beq}{\begin{equation}} 
\newcommand{\feq}{\end{equation}} 
\newcommand{\barr}{\begin{array}} 
\newcommand{\earr}{\end{array}}

\newcommand{\enums}[1]{\begin{enumerate} #1 \end{enumerate}}

\newcommand{\arry}[2]{\begin{array}{#1} #2 \end{array}}
 
\newcommand{\non}{\nonumber}

\newcommand{\Cplx}{\mathbb{C}}

\renewcommand{\ESO}{E_6/[SO(10)\times U(1)]}

\begin{document} 
\pagestyle{empty} 
\begin{flushright}
IC/2004/121\\
\end{flushright} 
\vs{6} 
\begin{center} 
{\large {\bf Phenomenological analysis of supersymmetric $\gs$--models on \\
\vs{2} 
coset spaces
{\boldmath $\bf SO(10)/U(5)$} and 
{\boldmath $\bf E_6/[SO(10)\times U(1)]$}}}\\
\vs{6} 

{\bf T.S.\ Nyawelo}\\
\vs{2} 
{\em The Abdus Salam International Centre for Theoretical Physics}\\ 
{\em Strada Costiera 11, I-34014 Trieste, Italy.}\\
\vs{1}
{\em E-mail: tnyawelo@ictp.trieste.it}\\ 
\vs{6}
{\small{ \bf{Abstract} }} \\
\end{center}

\nit
{\footnotesize{We discuss some phenomenological aspects of 
gauged supersymmetric $\gs$--models on homogeneous coset-spaces $\ESO$ 
and $SO(10)/U(5)$ which are some of the most interesting for 
phenomenology. We investigate in detail the vacuum configurations of these 
models, and study the resulting consequences for supersymmetry breaking 
and breaking of the internal symmetry. Some supersymmetric minima for both 
models with gauged full isometry groups $E_6$ and $SO(10)$  are 
physically problematic as the \Kh\ metric becomes singular and hence the kinetic terms of the Goldstone boson multiplets 
vanish. This leads us to introduce recently proposed soft supersymmetry-breaking mass 
terms which displace the minimum away from the singular point. A non-singular \Kh\ 
metric breaks the linear subgroup  $SO(10)\times U(1)$ of the $E_6$ model spontaneously. The particle spectrum of all these different models is computed.}}
\vfill 

\nit

\np 
~\hfill 

\np

\pagestyle{plain} 
\pagenumbering{arabic} 

\section{Introduction}
\nit
Non-linear supersymmetric $\gs$-models based on homogeneous \Kh ian cosets 
spaces $G/H$ may have applications to physics beyond the standard 
model. For example, supersymmetric extensions of a 
Grand Unified Theory (GUT) may be relevant for particle physics since they 
contain less parameters than the Minimal Supersymmetric Standard Model 
(MSSM). One of the original 
guidelines of the construction of GUT theories was renormalizability. 
However, as such GUT models are likely to be realized quite close to the 
Planck scale, renormalizability is not necessarily  an issue as 
supergravity theories are non-renormalizable by themselves. Moreover, supergravity models 
often include non-linear coset models such as $SU(1,1)/U(1)$ in $N = 4$.  
Therefore a GUT   
may be part of a supersymmetric non-linear sigma model 
based on a coset space $G/H$, with $H$ a subgroup of $G$.

For the construction of this kind of models the coset space $G/H$ must be a 
\Kh\ manifold  \ct{zumino,freed-alg}. The chiral fermion content of 
supersymmetric $\gs$-models based on 
homogeneous \Kh ian cosets spaces is often anomalous. The presence of chiral anomalies in internal symmetries restricts the usefulness of these models for 
phenomenological applications. Therefore, anomalies have to be removed to 
allow for gauging the internal symmetries. This is achieved \ct{SJ1} by coupling 
additional  chiral superfields (generically
called {\em matter} superfields) carrying representations of the coset space 
$G/H$.  An important question in the context of 
supersymmetric matter is how it can be coupled to supersymmetric 
$\gs$-models on \Kh\ manifolds without spoiling the (possibly non-linear) 
invariance of the original theory. This is required for the cancellation  
of anomalies as shown in \cite{jv}.  Using the general procedure of canceling  
anomalies by coupling additional chiral superfields, consistent 
supersymmetric $\gs$-models on coset spaces, including 
among others the grassmannian models on  
$SU(N+M)/[SU(N) \times SU(M) \times U(1)]$, the orthogonal unitary coset 
models on manifolds $SO(2N)/U(N)$, as well as models on exceptional cosets 
like $E_6/[SO(10) \times U(1)]$, have been studied in great detail 
\cite{SJ1,STJ,jv,JVSG}.  

Since $E_6$ and $SO(10)$ are promising unification groups, the coset spaces 
$E_6/[SO(10)\times U(1)]$ and $SO(10)/[SU(5)\times U(1)]$ are the most 
interesting for (direct) phenomenology. In the $E_6/[SO(10)\times U(1)]$ model, the 
fermion partners of the Goldstone bosons ---the quasi-Goldstone fermions--- 
have 
precisely the right quantum numbers to describe one family of quarks and 
leptons, including a right-handed neutrino. The model on $SO(10)/U(5)$ 
contains the $SU(5)\times U(1)$ fermionic field content of one generation of quarks and 
leptons, including a right-handed neutrino as well. Therefore, these models could have interesting phenomenological applications. In earlier studies of anomaly-free extension of 
supersymmetric $\gs$-model on $SO(10)/U(5)$, it was found that  
upon gauging the full $SO(10)$ the $D$-term potential sometimes force the 
scalar fields to take vacuum expectation values for which the model becomes 
singular, in the sense that the kinetic energy terms of the Goldstone boson and 
quasi-Goldstone fields disappear in the vacuum 
state, and the space of physical degrees of freedom is reduced. 
In a recent paper \cite{TsnFrJvSgn} we have investigated 
singularities in field geometry, where the kinetic terms vanish, by studying  a 
simple supersymmetric
model based on the homogeneous space $\Cplx P^1$. We  showed that the 
metric singularities can be regularized by addition of a soft 
supersymmetry-breaking mass parameter. 
 
The present paper is  a first step in the analysis of the phenomenology of those 
models. In order to discuss various properties of the models, we first review the 
construction of the lagrangians on coset-spaces that are 
globally consistent. We describe how anomaly cancellation can be achieved in 
supersymmetric $\gs$-model on $SO(10)/U(5)$ and $\ESO$, by adding matter 
fields; we then discuss the several interesting gauge extended versions of these models, and 
the resulting mass spectra.

This paper is structured as follows. The main aspects of  gauged  supersymmetric $\gs$--models on \Kh\ cosets with anomalies 
canceled  by matter fields is reviewed in section \ref{2}. In section \ref{3} we  derive the mass sum rule for non-linear supersymmetric $\gs$--models. These relations play an important role in constructing 
realistic supersymmetric gauge theories, containing the standard model.  In section 
\ref{4} we summarize the anomaly-free supersymmetric $\gs$--model on 
$SO(10)/U(5)$ as described in \cite{STJ}. We perform a quite 
general analysis of gauging the full $SO(10)$ group in subsection \ref{4.1}. We 
investigate in particular the existence of zeros of the potential, and show that  the models with fully gauged $SO(10)$ are singular. In subsection \ref{4.2} we extend the model with soft supersymmetry breaking mass terms which preserve the non-linear $SO(10)$. To complete the phenomenological analysis, 
we also consider the gauging of the linear subgroup $SU(5)\times U(1)$ of the $SO(10)$-spinor model in subsection \ref{4.3}. Because this subgroup 
contains an explicit $U(1)$ factor, we added a Fayet-Iliopoulos term with 
parameter $\gx$ and we investigate in particular the existence of zeros of the 
potential, for which the model is anomaly-free, with positive definite kinetic 
energy. Then we discuss a number of physical aspects of these models, like 
supersymmetry and internal symmetry breaking, and the resulting 
mass-spectrum. Section \ref{5} is devoted to phenomenological analysis of 
$\ESO$ model. We first summarize the results obtained in \cite{ysj1,YacSaoJv,SJ1}. 
Section \ref{c7s2} discusses the gauging of 
internal symmetries in general. In section \ref{c7s3}, we consider in some 
detail the gauging of the full non-linear $E_6$ symmetry. Like in the on 
$SO(10)/U(5)$, in one of the supersymmetric minima, we find 
that the $D$-term potential drives the scalar fields to a singular point of the 
kinetic terms. We show that the singular metric can also be regularized by the addition of a soft supersymmetry-breaking mass 
parameter. Gauging the linear subgroup $SO(10)\times U(1)$ gives consistent models, but only for special values of couplings constant and non-zero value of the Fayet-Iliopoulos term. Section \ref{6} contains the conclusions.

\section{Supersymmetric {\boldmath$\gs$}--models on 
\Kh\ manifolds \label{2}}

$N = 1$ globally supersymmetric 
lagrangians for non-linear $\gs$-models in 4-D space time, are formulated in terms of 
chiral superfields $\gF^\ga = (z^\ga, \gps^\ga_L, H^\ga )$, $\ga = 1,\dots,N$ 
the components of which are complex scalars $z^\ga$, an auxiliary 
field $H^\ga$ and a (left-handed) chiral fermion
\footnote{Our conventions for chiral spinors are such, that $\gg_5 \gps_L 
= +\gps_L$ and $\bgps_L \gg_5 = -\bgps_L$; charge conjugations acts as 
$\gps_R = C \bgps_L^T$, where $\bgps_L = 
i \gps_R^{\dagger} \gg_0$.} $\gps_L$. The action is defined by two functions of 
superfields: the real \Kh\ potential $K(\bgF,\gF)$, and the holomorphic 
superpotential $W(\gF)$. The component lagrangian after eliminating the auxiliary 
fields is \cite{zumino}
\begin{eqnarray}
\cL_{\mathrm{chiral}} &=& - G_{\ga \uga}(z, \bar{z}) \Bigl(
 \der^{\mu} z^{\ga}\, \der_{\mu} \bar{z}^{\,\uga} +
 \bar{\psi}_L^{\uga} \stackrel{\leftrightarrow}{\sDer}
 \psi_L^\ga\Bigl)  + R_{\ga\ugb,\gg\ugd}\,
\bar{\psi}^\ga_R\,\psi^\gg_L\bar{\psi}^\ugb_L\,
\psi^\ugd_R\nonumber\\
[2mm]
& & 
- G^{\ga \uga}\bW_{;\uga}W_{;\ga} + W_{;\ga\gb}(z)\bgps^\ga_R\gps^\gb_L + 
\bW_{;\uga\ugb}(\bz)\bgps^\uga_L\gps^\ugb_R.
\labl{chiral}
\end{eqnarray}
In this expression, we have used the following notation for the metric, connection and curvature constructed from the K\"{a}hler potential $K$, respectively:
\begin{eqnarray}
G_{\ga\uga} = K_{,\ga\uga},\quad\gG^\ga_{\gb\gg} = G^{\ga\uga}\,G_{\uga\gb,\gg},
\quad R_{\ga\ugb\gg\ugd} = G_{\ga\ugb,\gg\ugd} - G_{\ga\ugs,\gg}G^{\gs\ugs}G_{\ugb\gs,\ugd},
\end{eqnarray}
with $G^{\ga\uga}$ the inverse of the metric $G_{\ga\uga}$. The comma denotes 
differentiation with respect to $z^\ga$, $\bz^\uga$, while the semicolon denotes 
a covariant derivative. Moreover, the \Kh\ covariant derivative of a chiral 
spinor and the left-right arrow above the covariant derivative are 
defined by
\begin{eqnarray}
\sDer\gps^\ga_L = \sder\gps^\ga_L  + \gG_{\gb\gg}^\ga\sder z^\gb
\gps^\gg_L,\quad\bgps^\uga_L\stackrel{\leftrightarrow}{\sder}\gps^\ga_L = 
\bgps^\uga_L\gg^\gm\der_\gm\gps^\ga_L  - \der_\gm\bgps^\uga_L\gg^\gm\gps^\ga_L.
\end{eqnarray}

In general, the \Kh\ metric may admits a set of holomorphic isometries 
$R^\ga_i(z)$,  $\bR^\uga_i(\bz)$ $(i = 1,\dots, n)$, which are the solutions of the 
Killing equation
\begin{eqnarray}
R_{i\uga,\ga} + \bR_{i\ga,\uga} = 0.
\labl{kequ}
\end{eqnarray}
These isometries define infinitesimal symmetry transformations 
on the \Kh\ manifold $G/H$. In components the transformation rules read
\begin{eqnarray}
\gd z^\ga = \gth^i\,R^\ga_i(z),\quad\gd\bz^\uga =\gth^i\,\bR^\uga_i(\bz),\quad
\gd\psi^\ga_L = \gth^iR^\ga_{i,\gb}(z)\,\psi^\gb_L,\quad\gd\bgps^\uga_L = 
\gth^i\bR^\uga_{i,\ugb}(\bz)\,\bgps^\ugb_L,
\label{sigmais}
\end{eqnarray}
with $\gth^i$ the parameters of the infinitesimal transformations. As a result, the isometries form a Lie algebra:
\begin{eqnarray}
R^\gb_{[i}\, R^{\ga}_{j],\gb} = R^\gb_i\, R^{\ga}_{j,\gb} - R^\gb_j\, R^{\ga}_{i,\gb} = f_{ij}\,^k R^\ga_k.
\labl{liealgebra}
\end{eqnarray}
Thus, infinitesimal transformations (\ref{sigmais}) define a (generally non-linear) 
representation of some Lie group $G$, called the isometry group  of the manifold. 
The $f_{ij}\,^k$ are structure constants of the algebra. A special feature of \Kh\ manifolds is that the isometries can locally be written as the gradient of some real scalar functions, the  Killing potentials $M_i(z,\bz)$ \cite{bw,ysj1}:
\begin{eqnarray}
R^\ga_i = - i G^{\ga\uga} M_{i,\uga},\quad \bR^\uga_i = i G^{\ga\uga}M_{i,\ga}. 
\labl{kve}
\end{eqnarray}
From these equations, one sees that the Killing potentials $M_i$ are defined 
up to an integration constant $c_i$. It turns out that one can always choose 
these $c_i$ in such a way that the potentials $M_i$ transform in the adjoint 
representation of the isometry group:
\begin{eqnarray}
\gd_iM_j = R^\ga_iM_{j,\ga} + \bR^\uga_iM_{j,\uga} = -i 
G_{\ga\uga}\Bigl(R^\ga_i\bR^\uga_j - R^\ga_j\bR^\uga_i\Bigl)
 = f_{ij}\,^k M_k.
\end{eqnarray}
Under the transformation (\ref{sigmais}) the \Kh\ potential itself 
transforms as 
\begin{eqnarray}
\gd_i\,K = F_i(z) + \bF_i(\bz).
\labl{ktrans}
\end{eqnarray}
Now it can be shown that the functions $F^i$, $\bF^i$  defined by
\begin{eqnarray}
F_i = K_{,\ga}R^\ga_i + i M_i,\quad \bF_i = K_{,\uga}\bR^\uga_i - 
i M_i,
\labl{killingpot}
\end{eqnarray}
are holomorphic:
\begin{eqnarray}
F_{i,\uga} =0,\qquad \bF_{i,\ga} = 0.
\end{eqnarray}
From the Lie-algebra (\ref{liealgebra}) it follows that one can choose the 
transformations of the functions $F_i(Z)$ to have the property
\begin{eqnarray}
\gd_iF_j - \gd_jF_i = f_{ij}\,^kF_k 
\labl{1.10.2}.
\end{eqnarray}

We now turn to the possibility of realizing the transformation (\ref{sigmais}) locally. This is 
possible only if the symmetries are non-anomalous. 
As it is well known \cite{SJ1,sgn}, such anomalies 
can be removed by coupling additional chiral fermions $\gc^A_L$ contained in other 
chiral 
superfields $\gPs^A = (a^A, \gc^A_L)$ carrying specific
line-bundle representations of the group $G$.  Then the  complete superfield
content $\gS^I = (\gF^\ga, \gPs^A)$ of the model is specified by a scalar
superfield $\gF^\ga = (z^\ga, \psi^\ga_L, H^\ga)$, which includes the complex
coordinate $z^\ga$ of this manifold $G/H$, and a set of matter superfields $\gPs^A = 
(a^A, \gc^A_L, F^A)$.

Once the anomalies have been canceled, 
the $G$ symmetry group can be gauged in a way that respects the supersymmetry. In summary, one first introduces a set of vector multiplets 
$V^i = (A^i_\gm, \gl^i_L, D^i)$, where $A^i_\gm$ is a 
gauge field, $\gl^i_L$ a gaugino and $D^i$ is an auxiliary complex scalar. 
This gives rise to introduction  of the  gauge covariant derivatives, accompanied by Yukawa and a 
D-term potential, defined in terms of the Killing potential $M(z,\bz)$ for the 
isometries group $G$. And  finally, one introduces the kinetic terms for the vector 
multiplets. Then the full  lagrangian of globally anomaly-free supersymmetric 
$\gs$-models on \Kh\ manifolds, after eliminating the auxiliary fields 
$(H^\ga, F^A, D^i)$ becomes
\begin{eqnarray}
\cL &=& - G_{I\uI}\Bigl(D\bZ^\uI\cdot D 
Z^I + \bar{\psi}^I_L\stackrel{
\leftrightarrow}{D\Slashed}\psi_L^\uI\Bigl) 
- \Bigl(G_{I\uI,I}D_\gm Z^J - G_{I\uI,\uJ}D_\gm\bZ^\uJ\Bigl)
\bgps^\uI_L\gg_\gm\gps^I_L \nonumber\\
[2mm]
& & + 
R_{I\uI J\uJ}\,\bar{\psi}^I_R\,\psi^J_L\bar{\psi}^\uI_L\,
\psi^\uJ_R - \frac{g^2}{2}(\cM_i + \gx_i)^2
+ 2 \sqrt{2}\,g\,G_{I\uI}\Bigl(\bR^\uI_i\bgl^i_R\gps^I_R + 
R^I_i\bgl^i_L\gps^\uI_L\Bigl)\nonumber\\
[2mm]
& & 
- \frac{1}{4}F^i\cdot F^i - 
\bgl^i_R\stackrel{\leftrightarrow}{D\Slashed}\gl^i_R.
+ \bW_{;\uI\uJ}\bgps^\uI_L\,\psi^\uJ_R + 
W_{;I J}\bgps^I_R\,\psi^J_L - G^{I\uI}\,\bW_{;\uI}\,W_{;I}.
\labl{generallag}
\end{eqnarray}
Here we have added a Fayet-Iliopoulos term with parameter $\gx_i$ in 
case there is a commuting $U(1)$ vector multiplet and $\cM_i(z,\bz;a,\ba)$ is an extended version of the Killing potentials introduced in (\ref{kve}). Furthermore, the notation $Z^I = (z^\ga, a^A)$ and $\gps^I_L = (\gps^\ga_L, \gc^A_L)$, $I = 
(\ga,A)$,  denote the scalar and spinor components of the superfields $\gS^I$. 
The covariant derivatives  contained the gauge fields and field strength tensor 
$F^i_{\gm\gn}$ are
\begin{eqnarray}
D_\gm Z^I &=& \der_\gm Z^I - gA^i_\gm R^I_i,\quad\quad\,
D_\gm\gps^I_L = \der_\gm\gps^I_L - g\,A^i_\gm R^I_{i,J}\gps^J_L,\nonumber\\
[2mm]
D_\gm\gl^i_R &=& \der_\gm\gl^i_R - g\,f^{ijk}\,A^j_\gm\,\gl^k_R,\quad 
F^i_{\gm\gn} = \der_\gm\,A^i_\gn - \der_\gm\,A^i_\gm - g\,f^{ijk}\,A^j_\gm\,A^k_\gn. 
\labl{GCovariant}
\end{eqnarray}
 
\section{The mass formula\label{3}}

A very particular feature of a supersymmetric theories is the existence of a 
mass formula valid for all possible vacua with spontaneously broken 
supersymmetry and vacua preserving supersymmetry, relating the masses of 
all the fields present in 
the theory. This mass formula is very convenient when discussing realistic 
models. It is well known that a mass formula holds when supersymmetry is not 
broken: all states belonging to a given supermultiplet have the same mass. 
This result has for consequence the following sum rule. The supertrace of 
the mass matrices squared of all states: 
\begin{eqnarray}
{\rm STr}\, \rm{m}^2 = \Tr\Bigl(\rm{m}^2_0 + 3 m^2_1 - 2 m^2_{\frac{1}{2}}\Bigl),
\labl{sumrule}
\end{eqnarray}
where $\mathrm{m}^2_1$, $\mathrm{m}^2_{\frac{1}{2}}$ and $\mathrm{m}^2_0$ are respectively the mass 
matrices squared of spin--1, $\frac{1}{2}$ (four component spinors) 
and $0$ (real scalars) states of the theory. For a supersymmetric 
multiplet of a mass $\mathrm{m}$, ${\rm STr}\,\mathrm{m}^2$ is defined so that
\begin{eqnarray}
{\rm STr}\,\mathrm{m}^2 = \sum\mathrm{m}^2\,(\textrm{number of bosons} -  \textrm{number of fermions}) 
= 0.
\end{eqnarray}
However, the vanishing of the supertrace for a supersymmetric theory is much 
weaker than statement of the equality of all masses within a supermultiplet. 
Indeed a formula for ${\rm STr}\,\rm{m}^2$ can be generalized to arbitrary 
vacua, including those breaking supersymmetry \cite{eslp}. 
The standard choice for vacuum configurations is to allow for constant 
values of Lorentz 
invariant fields. Thus only scalars $Z^I$ are allowed to have a non-zero 
vacuum expectation values (v.e.v.), denoted by $\langle Z^I\rangle$. For 
this configuration, the theory reduces to the scalar potential 
\begin{eqnarray}
V = - \cL\Bigl(\der_\gm \langle Z^I\rangle = \langle \gps^I_L\rangle = 
\langle 
\gl^i_L\rangle = \langle A^i_\gm \rangle = 0\Bigl).
\end{eqnarray}
In this section we derive the supertrace formula for supersymmetric 
non-linear $\gs$-models described by (\ref{generallag}) relevant for later 
applications. Since in the models we consider in this paper, the isometry 
group $G$ does
not allow for an invariant trilinear superpotential $W(\gS)$, we will not consider  
here the contributions of $W(\gS)$ to the mass formula. For this reason, 
from now on 
we take $W(\gS) = 0$ (hence the terms involving $W(Z)$ in the full lagrangian 
(\ref{generallag}) are absent). In order to calculate ${\rm STr}\,\rm{m}^2$, 
we need the explicit form of the three mass matrices in (\ref{sumrule}). 

We first consider the mass matrix squared for a spin--1 particle.  
When the scalar fields $Z^I$ acquire a vacuum expectation value, some 
gauge bosons will become massive in general. From (\ref{generallag}), 
the part of the lagrangian quadratic in spin--1 particles is
\begin{eqnarray}
\cL_1 =  - G_{I\uI} \,D\bZ^\uI\,\cdot D Z^I
- \frac{1}{4} F^i\cdot F^i,
\labl{lag1}
\end{eqnarray}
with the field strength and the covariant derivative defined in (\ref{GCovariant}). Substituting the 
expressions for the field strength and the covariant 
derivative the lagrangian (\ref{lag1}) becomes:
\begin{eqnarray}
\cL_1 = - \frac{1}{2}\Bigl[(\der_\gm A^i_\gn)(\der^\gm A^{i\gn} - 
\der^\gn A^{i\gm})  - 2 g^2 \langle R_i^I\bR^\uI_j\,G_{I\uI}
\rangle A_\gm^i A^{j\gm}\Bigl].
\labl{spin}
\end{eqnarray}
This expression means that the mass matrix (squared) of spin--1 particles is
\begin{eqnarray}
(\mathrm{m}^2_1)_{ij} = 2 g^2 \langle R_i^I\bR^\uI_j\,G_{I\uI}
\rangle
\labl{mass2}
\end{eqnarray}
From (\ref{mass2}), the trace of the mass matrix squared for gauge fields 
$A^i_\gm$ is
\begin{eqnarray}
3\Tr\,\mathrm{m}^2_1 = 6g^2 \langle R_i^I\bR^\uI_i\,G_{I\uI}
\rangle = 6 g^2\,\langle G^{I\uI}\cM_{i,I}\cM_{i,\uI}\rangle.
\labl{mass1}
\end{eqnarray}
The last equality follows up on using (\ref{kve}). 

Turning to the spin--$\frac{1}{2}$ mass matrix, we collect all the terms bilinear 
in fermionic fields in lagrangian (\ref{generallag}) with possible vacuum expectation values $\langle Z^I\rangle$. They read
\begin{eqnarray}
\cL_{\frac{1}{2}} = - 2 \langle G_{\uI I}\rangle
\bgps^\uI_L\,\der\slashed\,\psi_L^{I} - 2\bar{\lambda}^i_L\der\slashed\lambda^i_L +
2 \sqrt{2}i\, g \Bigl(\cM_{i,I}\bgl^i_R\,\psi^I_L  - 
\cM_{i,\uI}\bgps^\uI_L\gl^i_R\Bigl)  
+ \dots
\labl{massterm1},
\end{eqnarray}
where the dots represent total derivatives terms that do 
not affect the action.  The non-vanishing mass term 
can be  written in a matrix form as
\begin{eqnarray}
\cL_{\frac{1}{2}} = - 2 \langle G_{I\uI}\rangle
\bgps^\uI_L\,\der\slashed\,\psi_L^{I} + 2\bar{\lambda}^i_L\der\slashed\lambda^i_L + 2 \left(
\begin{array}{cc}
\bgps^\uI_L&\bar{\lambda}_R^i
\end{array}
\right)M_F\left(
\begin{array}{c}
\gps^I_L\\
\lambda^j_R
\end{array}
\right)
\labl{FermionMassess2}
\end{eqnarray}
with the fermion mass matrix evaluated at the classical minimum of the 
potential
\begin{eqnarray}
M_F = \left(
\begin{array}{cc}
0&-i\sqrt{2}g\,\cM_{i,\uI}\\
i\sqrt{2}g\,\cM_{i,I}&0
\end{array}
\right).
\labl{MF}
\end{eqnarray}
From this expression we obtain the mass matrix squared of spin--$\frac{1}{2}$ 
particles 
\begin{eqnarray}
\Bigl(M^2_{\frac{1}{2}}\Bigl) = \Bigl(M_F\,M^\dag_F\Bigl) = \left(
\begin{array}{cc}
2g^2\,\langle\cM_{i,\uI}\cM_{i,I}\rangle&0\\
0&2g^2\,\langle G^{I\uI}\,\cM_{i,\uI}\cM_{j,I}\rangle
\end{array}
\right).
\labl{Msquared}
\end{eqnarray}
This mass matrix has to be normalized such that the kinetic terms of the 
fermionic fields take the standard form
\begin{eqnarray}
\cL_{\mathrm{Dirac}} = - 2 \bgch^I  (\der\slashed - M_{IJ} )\gch^J.
\labl{dmass}
\end{eqnarray}
This is achieved by multiplying the mass matrix (\ref{MF}) with the inverse 
metric $G^{I\uI}$ and introduces the Dirac fermions as a 
combination of a 
left-handed chiral fermions $\gps^I_L$ and the right-handed 
gauginos $\gl^i_R$.
As a result, the trace of the mass matrix squared of spin--$\frac{1}{2}$ particles is then
\begin{eqnarray}
\Tr\,\mathrm{m}^2_{\frac{1}{2}} = \Tr\,M^2_{\frac{1}{2}} = 
4 g^2\,\langle G^{I\underline{I}}
\cM_{i,I}\cM_{i,\uI}\rangle.
\labl{mass1/2}
\end{eqnarray}

The last thing we need is the scalar mass matrix (squared).  
The lagrangian has the form
\begin{eqnarray}
\cL_0 = - G_{\uI I}\,\der Z^I\cdot
\der\bZ^\uI - V(Z,\bZ).
\end{eqnarray}
By expanding the scalar potential $V(Z,\bZ)$ to second order in 
complex fluctuation $\tZ^I$ around the minimum 
$Z^I = \langle Z^I\rangle$, the bilinear terms are 
\begin{eqnarray}
\cL_0 &=& - \langle G_{I\uI}\rangle\,\der\tZ^I\cdot
\der\tilde{\bZ}^\uI + \langle V_{,\uI I}
\rangle\tZ^I\,\tilde{\bZ}^\uI + 
\frac{1}{2}\langle V_{,I J}
\rangle\tZ^I\tZ^J + \frac{1}{2}\langle V_{,\uI\uJ}
\rangle\tilde{\bZ}^\uI\tilde{\bZ}^\uJ\nonumber\\
[2mm]
&=& - \langle G_{I\uI}\rangle\,\der\tZ^I\cdot
\der\tilde{\bZ}^\uI -\frac{1}{2}\left(
\begin{array}{cc}
\tZ^I & \tilde{\bZ}^\uI
\end{array}
\right)\,M_0^2
\left(
\begin{array}{c}
\tilde{\bZ}^\uJ\\
\tZ^J
\end{array}
\right),
\end{eqnarray}
with the spin--$0$ mass matrix squared $M_0^2$:
\begin{eqnarray}
M_0^2 = \left(
\begin{array}{cc}
\langle V_{I\uJ}\rangle & \langle V_{IJ}\rangle \\
[2mm]
\langle V_{\uI\uJ}\rangle & 
\langle V_{\uI J}\rangle
\end{array}
\right).
\labl{smass}
\end{eqnarray}
In a similar fashion the bosonic mass eigenstates have to be normalized 
such that their kinetic lagrangian takes the standard form. This is achieved again by multiplying the mass matrix squared 
(\ref{smass}) with the inverse metric 
$G^{\uI I}$:
\begin{eqnarray}
\Tr\,\tM^2_0 =  2\,\langle G^{I\uI}\,V_{\uI I}\rangle.
\labl{trac}
\end{eqnarray}
From the scalar potential
\begin{eqnarray}
V =  \frac{g^2}{2}(\cM_i + \gx_i)^2 
\end{eqnarray}
obtained from our general lagrangian (\ref{generallag}), one has
\begin{eqnarray}
V_{\uI I} = g^2\Bigl(\cM_{i\uI}\,\cM_{iI} + 
(\cM_i + \gx_i)\,\cM_{i\uI I}\Bigl).
\labl{VAA} 
\end{eqnarray}
After substituting the second mixed derivative of the scalar 
potential (\ref{VAA}) in (\ref{trac}) we obtain the trace 
of the spin-less mass matrix squared:
\begin{eqnarray}
\Tr\,\mathrm{m}^2_0 &=& 2\,g^2\,G^{\uI I}\Bigl(\cM_{i\uI}\,\cM_{iI} + 
(\cM_i + \gx_i)\,\cM_{i\uI I}\Bigl).
\labl{mass0}
\end{eqnarray}
Finally, collecting results (\ref{mass0}), (\ref{mass1}) and 
(\ref{mass1/2}) leads to the general mass sum rule for 
non-abelian gauged supersymmetric non-linear sigma models 
without a superpotential:
\begin{eqnarray}
{\rm STr}\,\mathrm{m}^2 &=&  2 g^2\,G^{\uI I}\,(\cM_i + \gx_i)\,\cM_{i\uI I},
\quad D^i = (\cM_i + \gx_i)
\labl{sumrule1}
\end{eqnarray} 
which is valid for arbitrary vacuum expectation values 
$\langle Z^I \rangle$.

The general mass sum rule for 
Yang-Mills theories with local supersymmetry, was derived by Cremmer, 
Ferrara, Girardello and van Proeyen \cite{eslp}. It has also been derived 
in superspace by considering  1-loop divergences 
\cite{Grisaru,Gates,GiraRockKar} 
in the (non--singular) field space 
\begin{eqnarray}
{\rm STr}\,  \mathrm{m}^2 = 2 i D_i R^{iI}\,_{;I} = 
2 i D_i\Bigl[R^{iI}\,_{, I} + R^{iJ}\,\gG^I_{JI}\Bigl].
\labl{CG}
\end{eqnarray}
The equivalence of this result (\ref{CG}) to ours (\ref{sumrule1}) is 
rather easy to show using (\ref{kve}). 
Observe here, that the first term $R^{iI}\,_{, I}$ in (\ref{CG}) always vanishes in supersymmetric $\gs$-models on K\"{a}hler cosets
with anomalies canceled by matter as in models considered here (non-abelian gauged 
supersymmetric non-linear sigma models.)

Some comments are in order here about the formula 
(\ref{sumrule1}). It has been derived on the assumption 
that the \Kh\ metric $G_{\uI I}$ is invertible. However, in 
some cases as we will discuss in the following sections, the \Kh\ metric $G_{\uI I}$ develops a zero mode in the 
minimum of the potential; and the analysis of the theory becomes 
complicated by the appearance of the infinities at the classical level. 
A particular solution to this problem is to shift the minimum of potential 
away from the  position where the singularities occur by adding to 
the model extra terms which break supersymmetry 
explicitly. These new terms, which break supersymmetry without 
generating unwanted quadratic divergences are called soft breaking 
terms. 

Explicit breaking of global supersymmetry has been discussed in 
\cite{GiraGris}. Here we only focus on the scalar soft 
breaking mass term, relevant for later applications:
\begin{eqnarray}
\cL_{\mathrm{break}} =  |\gm|^2 X(Z,\bZ).
\labl{softbrea}
\end{eqnarray}
Here $X$ is real scalar which is invariant under the full set of the 
isometries $G$, and $\gm^2$ is real and nonzero.

After the addition of the soft breaking terms (\ref{softbrea}), the supertrace formula 
becomes
\begin{eqnarray}
{\rm STr}\, {\rm m}^2 &=&  2 g^2\,G^{\uI I}\,(\cM_i + \gx_i)\,\cM_{i\uI I} +
2 \gm^2\,G^{\uI I}\,X_{\uI I}.
\labl{sofsum}
\end{eqnarray}

\section{Analysis of particle spectrum of $\bf{SO(10)/U(5)}$--spinor model}
\label{4}

From the point of view of unification the coset space $SO(10)/[SU(5)\times U(1)]$ is a 
very interesting for  phenomenological applications as both $SO(10)$ 
and $SU(5)$ are often used GUT groups. However, a supersymmetric model built on the  
$SO(10)/[SU(5)\times U(1)]$ coset is 
not free of anomalies by itself as all the $\underline{10}$ anti-symmetric 
complex coordinates $z^{ij}$ and their chiral superpartners $\gps^{ij}_L$ ($i, j = 1,\dots, 5)$ of this manifold carry the same  
charges. To construct a consistent supersymmetric model on this coset one has to 
include the fermion partners of the  coordinates in an anomaly-free 
representation.  As $SU(5)$ representations are not anomaly free by themselves, 
we have to use the full $SO(10)$ representations for our additional matter 
coupling in this case. This has been achieved in \cite{STJ} by introducing 
a singlet $\underline{1}$ and completely anti-symmetric tensor with 4 indices 
which is equivalent to $\underline{\bar{5}}$ to complete the set of complex chiral 
superfields to form a $\underline{16}$ of $SO(10)$. The anti-symmetric 
coordinates of the coset are combined into a $\underline{10}$ of $SU(5)$ 
with a unit $U(1)$ charge. An anomaly free representation is obtained using the 
branching of the $\underline{16}$. Indeed, its decomposition under $SU(5)$ reads
\begin{eqnarray}   
\underline{16} = \underline{10}(1) + \underline{\bar{5}}(-3) + 
\underline{1}(5),
\end{eqnarray}   
where the numbers in parentheses denote the relative $U(1)$ charges. Therefore, the supersymmetric model on the coste $SO(10)/U(5)$ is defined by three chiral superfields 
$(\gF^{ij},\gPs_i,\gPs)$: the target manifold $SO(10)/U(5)$ is parametrized by 10 
anti-symmetric complex fields $z^{ij}$ in a chiral 
superfield $\gF^{ij} = (z^{ij}, \gps^{ij}_L, H^{ij})$, to which are added $SU(5)$ 
vector and scalar matter multiplets denoted respectively 
by $\gPs_i = (k_i,\go_{L~i}, B_i)$, and $\gPs = (h,\varphi_L, F)$. 

The complete \Kh\ potential of the model is
\begin{eqnarray}
\cK(z, \bz; k, \bk; h, \bh) &=& \frac{1}{2f^2}K_\gs(z, \bz) + K_{\underline{1}} + 
K_{ \underline{\bar{5}}}
,\nonumber\\
[2mm]
&=& \frac{1}{2f^2}\ln\det\gc^{-1} + 
|h|^2\,e^{-2 f^2 K_\gs} + e^{f^2 K_\gs}k\gc^{-1}\bk
\labl{comkh}
\end{eqnarray}
with the submetric $\gc^{-1} = \Id + f^2 z \bz$ and
$
\smash{e^{f^2 K_\gs} = (\det \gc)^{-1}}.
$ The dimensionfull constant $f$ is introduced to assign correct physical dimensions 
to the scalar fields $(z,\bz)$. The \Kh\ metric $G_{I \uI}$ derived from this 
\Kh\ potential $\cK$ possesses a set of holomorphic Killing vectors generating a 
non-linear representation of $SO(10)$:
\begin{eqnarray}
\gd z &=& \frac 1f\,  x - u^Tz - zu + f\,  zx^\dag z, 
\nonumber\\
[2mm] 
\gd  h &=& 2\tr(f\, zx^\dag - u^T) h,\nonumber\\
[2mm] 
\gd k &=&  - k \Bigl(-u^T + f\, zx^\dag  + 
\tr (-u^T + f\, zx^\dag) \Id\Bigl),
\labl{vkilling}
\end{eqnarray}
Here $u$ represents the parameters of the linear diagonal $U(5)$ transformations, 
and $(x, x^\dag )$ are 
the complex parameters of the broken off-diagonal $SO(10)$ transformations. It is readily checked that under the transformations (\ref{vkilling}) the \Kh\ potential $\cK$ 
transforms as in eq.~(\ref{ktrans}):
\begin{eqnarray}   
\gd\cK = \tr(f\, zx^\dag - u^T) + \textrm{h.c.} = F(z) + \textrm{h.c.}\, .
\end{eqnarray}   
This result guarantees the invariance of the metric, as expected if the the 
transformations (\ref{vkilling}) are isometries. Equivalently, one may check that the 
Killing vectors (\ref{vkilling}) satisfy  the Killing equation (\ref{kequ}) 
with  a metric of the form  
\begin{eqnarray}   
G_{I\uI}= \dsp{ \frac{\der^2 \cK}{\der Z^I \der\bZ^{\uI}} }
= 
\left(
\begin{array}{ccc}
{G_\gs}_{z^{ij}\bz_{kl}}
&  G_{z^{ij}\bk^i}&G_{z^{ij}\bh}
\\[2ex] 
G_{k_i\bz_{ij}} &  G_{k_i\bk^j}&0
\\[2ex] 
G_{h\bz_{ij}}& 0 &G_{h\bh}
\end{array}
\right).
\labl{kmetric}
\end{eqnarray}   

\subsection{Gauging of the full {\boldmath $\bf SO(10)$} isometries \label{4.1}}

In order for the chiral fermions $(\gps^{ij}_L,\go_{Li},\gvf_L)$ to have a 
physical interpretation as describing a family quarks and 
leptons, in 
this section we introduce gauge interactions. In this case supersymmetry 
implies the 
addition of a potential from  elimination of the auxiliary $D^i$ fields by
substitution for the Killing potentials \cite{WessBagger}.  We
consider the case in which  the full $SO(10)$ isometry is gauged. We
denote collectively the $SO(10)$ gauge fields as $A_\gm = (U_\gm,W^{\dag}_\gm, W_\gm)$  with $W^{\dag}_\gm$ and $W_\gm$ the gauge fields
corresponding to the broken $SO(10)$ transformations parametrized by 
$(x, x^\dag)$ and with $U_{\mu}$, the gauge field of the diagonal 
transformations parametrized by $u$.  This requires the introduction of covariant 
derivatives for the dynamical fields:
\begin{eqnarray}
D_\gm z &=& \der_\gm z - g_{10}\Bigl(\frac{1}{f} W_\gm - U^T_\gm z - 
U_\gm z 
+ f z W_\gm{}^\dag z\Bigl)
,\nonumber\\
D_\gm k &=& \der_\gm k + g_{10}\,k\,\Bigl(f W_\gm{}^\dag z - U^T_\gm
+ \tr(f W_\gm{}^\dag z - U^T_\gm )\Id\Bigl)
,\nonumber\\
D_\gm h &=& \der_\gm h - 2 g_{10}\tr\Bigl(f W_\gm{}^\dag z - U^T_\gm\Bigl)h.
\labl{CovDer} 
\end{eqnarray}
In the construction of these covariant derivatives we replaced the infinitesimal parameters $(x, x^\dag)$ by gauge fields. 

For the $D$--term scalar potential we need the $SO(10)$ Killing
potentials. The full Killing potential $\cM$ generating the Killing vectors  
(\ref{vkilling}) can be written as
\begin{eqnarray}
\cM(u,x^\dag, x) = \tr\Bigl(u\cM_u + x^\dag\cM_x + x\cM_x^\dag\Bigl),
\end{eqnarray}
with the $U(5)$ Killing potentials $\cM_u$, and the broken Killing 
potentials ($\cM_x,\cM_{x^\dag}$) given by \cite{STJ}
\begin{eqnarray}
-i\cM_u &=& M\,(\Id - 2f^2 \bz\gch z )
+ e^{f^2 K_\gs}(k^T\bk^T - f^2 \bz\bk k z),
\non \\[2mm]
-i\cM_{x^\dag} & = &
f \bz\gch\,M
+ f e^{f^2 K_\gs}\bz\bk k,
\labl{so10killing} \\[2mm]
-i\cM_x & =& - f \gch z\,M
- f e^{f^2 K_\gs}\bk k z,\quad M = \frac1{2f^2} -2 |h|^2\,e^{-2 f^2 K_\gs} + 
e^{f^2 K_\gs}k\gc^{-1}\bk.
\non 
\labl{Kilpot}
\end{eqnarray}
Alternatively, the $D$--term potential arising from gauging of $SU(5)\times U(1)$ including a 
Fayet-Iliopoulos parameter $\gx$ is
\begin{eqnarray}
V &=& \frac {g_1^2}{10} (\gx -i\cM_Y)^2 + \frac{g_5^2}{2} \tr (-i\cM_t)^2,
\labl{SU5xU1}
\end{eqnarray}
with $g_1$ and $g_5$ are the $U(1)$ and $SU(5)$  gauge couplings 
respectively. The $U(1)$ Killing potential $\cM_Y$ is defined as the trace of 
$U(5)$ Killing potential $\cM_u$ whereas the remaining $SU(5)$ 
Killing potential $\cM_t$ is defined as a traceless part of $\cM_u$: 
\begin{eqnarray}
\cM_t = \cM_u - \frac 15 \cM_Y \, \Id, 
 \qquad \cM_Y = \tr \cM_u.
\end{eqnarray}
The case of fully gauged $SO(10)$ is obtained by taking the coupling 
constants equal: $g_1 = g_5 = g_{10}$, and the 
Fayet-Iliopoulos term to vanish: $\gx = 0$. 

The coupling of the gauge multiplets to the supersymmetric non-linear $\gs$--model on 
$SO(10)/U(5)$ has interesting consequences for the spectrum. It can induce 
spontaneous breaking of  supersymmetry, and further spontaneous breaking of the 
internal symmetry. For example, if we gauge the full $SO(10)$, all the Goldstone 
bosons $(z,\bz)$ are absorbed by the vector bosons $(W^{\dag}_\gm, W_\gm)$ which 
become massive. In this case we may choose to study the model in the unitary 
gauge $ z = \bz = 0$. However, it was found in \cite{STJ} that in this gauge, the \Kh\ metric (\ref{kmetric}) develop zero-modes in the vacuum: the metric ${G_\gs}_{z^{ij}\bz_{kl}}$ for the Goldstone bosons and their fermions vanishes. 

To see this, we start from the scalar potential (\ref{SU5xU1}).  As already 
stated, we choose the unitary gauge: $z = \bz = 0$, $\gx = 0$, and $g_1 = g_5 = g_{10}$. Then  
the potential for the fully gauged $SO(10)$  model becomes
\begin{eqnarray}
V_{\mathrm{uni}} &=& \frac{g_{10}^2}{10}\Bigl(10|h|^2 - \frac{5}{2f^2} - 
6|k|^2\Bigl)^2 + 
\frac{2}{5}g_{10}^2\Bigl(|k|^2\Bigl)^2.
\labl{uni}
\end{eqnarray}
From this we see that we only have a supersymmetric minimum if 
\begin{eqnarray}
|k|^2 = 0,\quad |h|^2 = \frac{1}{4f^2}.
\end{eqnarray}
It can be seen immediately that this solution yields the vanishing of the 
\Kh\ metric:
\begin{eqnarray}
G_{z\bz} = {G_\gs}_{(ij)}{}^{(kl)} = \gd_{i}^{[k}\gd_{j}^{l]}\Bigl(\frac{1}{2f^2} - 
2|h|^2 + 
|k_i|^2\Bigl) + k^{(k}\gd_{(i}{}^{l)}\bk_{j)} = 0.
\labl{vanishing}
\end{eqnarray}
In this case the kinetic terms of the Goldstone superfield components vanish, 
therefore, mass terms for the $SO(10)$ gauge fields  $(W^{\dag}_\gm, W_\gm)$ 
vanish as well. Moreover, the theory becomes strongly 
coupled, with some of the four--fermion interactions exploding, namely:  
\begin{eqnarray}
\cL_{\mathrm{4-ferm}} = R_{z \bz h \bh} \,  \bgps_R \gps_L \,  \bgvf_L \gvf_R 
+ {\rm perm.}
\labl{4fermi} 
\end{eqnarray}
with the curvature components given by  
\begin{eqnarray}
\barr{lll}
R_{z \bz h \bh} =  {R}_{(ij)}{}^{(kl)}{}_{h\bh} = -2 f^2\gd_{i}^{[k}\gd_{j}^{l]}
\Bigl(1 +  2|h|^2(\frac{1}{2f^2} - 2|h|^2 )^{-1}\Bigl).
\earr
\label{4fermicouplings}
\end{eqnarray}
This may point to a restauration of the $SO(10)$ symmetry. Clearly, not all of 
the physics described by this model is yet understood.  

\subsection{Softly broken supersymmetry\label{4.2}}

To avoid the problem of vanishing of the 
\Kh\ metric, we shift the minimum of the 
potential away from the singular point by adding $SO(10)$-invariant soft 
supersymmetry breaking scalars mass terms
\begin{eqnarray}
\gD V =  \mu_1^2\,|h|^2\,e^{-2 f^2 K_\gs} + \mu_2^2\,e^{f^2 K_\gs}k\gc^{-1}\bk
\labl{so10soft}
\end{eqnarray}
to the potential. As a result the minimum of the potential is shifted to a 
position where the 
expectation value of the K\"{a}hler metric is not
vanishing; and the scalar $h$ gets a 
vacuum expectation value 
\begin{eqnarray}
|k|^2 = 0,\quad |h|^2 = v^2 = \frac{1}{4f^2} - \frac{\gm_1^2}{20g_{10}^2},\quad \gm_1^2 < \frac{5g^2_{10}}{f^2},
\labl{MINIMUM}
\end{eqnarray}
breaking the linear local $U(1)$ subgroup. The corresponding $U(1)$ vector becomes massive; 
and the remaining vectors of $SU(5)$ stay 
massless. In the fermionic sector, two Dirac fermions are  realized as a 
combination of the fermions of the chiral multiplets with the gauginos.

We now present details of the above mass spectrum. Since in general $SO(10)$ is 
broken in the vacuum, the Goldstone bosons $(\bz, z)$ are absorbed in
the longitudinal component of the charged vector bosons, and 
we may choose the unitary gauge $\bz = z = 0$. 
In this gauge the  K\"{a}hler metric in the minimum (\ref{MINIMUM}) 
is automatically diagonal:
\begin{eqnarray}
G_{I\uI} = \left(
\begin{array}{ccc}
{G_\gs}_{(ij)}{}^{(kl)}&0&0\\
0&G^i_j&0\\
0&0&G_{h\bh}
\end{array}
\right) =\left(
\begin{array}{ccc}
\gd_{i}^{[k}\gd_{j}^{l]}\,\frac{\gm^2_1}{10g^2_{10}}
&0&0\\
0&\gd^i\,_j&0\\
0&0&1
\end{array}
\right),
\labl{Kahler metric}
\end{eqnarray}
and all the $z$ dependence is removed from the covariant derivatives 
(\ref{CovDer}). To calculate the bosonic mass spectrum, we consider the bosonic part of 
the model, which up to the kinetic terms for the 
gauge bosons is described by the action
\begin{eqnarray}
\cL_{\mathrm{bos}} &=&  
- g^2_{10}\,{G_\gs}_{(ij)}{}^{(kl)}\bW^{(ij)}\cdot W_{(kl)} - D\bk^i\cdot D k_i - 
D\bh\cdot D h - V_{\mathrm{full}}\nonumber\\
[2mm]
& &
- \frac{1}{4}\Bigl[\frac{1}{2}\bF_{(ij)}(W)\cdot F^{(ij)}(W) + 
F^i\,_{j}(U)\cdot F^i\,_{j}(U)\Bigl] + \dots,
\labl{bos}
\end{eqnarray}
where $V_{\mathrm{full}}$ is given by $V_{\mathrm{uni}}$ eq.~(\ref{uni}) and $\gD V$ 
eq.~(\ref{so10soft}). In this expression the covariant derivatives include 
only the $U(5)$
gauge field. To identify the masses of  the gauge fields,
we decompose the $U(5)$ vector multiplet $U^i_{~j} = (U^i_{\gm j},\gL^i_{Rj})$ 
into a $U(1)$  and $SU(5)$ vector multiplets denoted respectively by 
$A = (A_\gm, \lambda_R)$  
and $V^i_{~j} = (V^i_{\gm j},\gl^i_{Rj})$:
\begin{eqnarray}
V = U - \frac{1}{5}A\Id_5\quad \tr (V) = 0,\quad A = \tr (U).
\labl{U5}
\end{eqnarray}
It follows that the kinetic terms for the $SU(5)\times U(1)$ 
gauge fields become
\begin{eqnarray}
- \frac{1}{4}\tr F_{\gm\gn}^2(U)  - \bgL^j_{Ri}\stackrel{\leftrightarrow}{\sder}
\gL^i_{Rj} &=&  - \frac{1}{5}\Bigl(\frac{1}{4}F_{\gm\gn}^2(A) + \bgl_R\stackrel{\leftrightarrow}{\sder}{\gl}_{R}\Bigl) + 
\frac{1}{4}\tr [F_{\gm\gn}^2(V)]
\nonumber\\
[2mm]
& &
- \bgl^j_{Ri}\stackrel{\leftrightarrow}{\sder}\gl^i_{Rj}.
\end{eqnarray}
Notice that the kinetic terms for the $U(1)$ multiplet are not canonically normalized.  
To obtain the standard normalization,  we redefine the $U(1)$ multiplet according to
\begin{eqnarray}
A\rightarrow\sqrt{5}(\tilde{A_\gm}, \tilde{\lambda}_R).
\end{eqnarray}
With the redefined fields, the kinetic terms for the gauge fields become
\begin{eqnarray}
\cL_{\mathrm{gauge}} &=& - \frac{1}{4}\Bigl[\frac{1}{2}\bF_{(ij)}(W)\cdot F^{(ij)}(W) +
F_{\gm\gn}^2(\tilde{A}) + F^i\,_{j}(V)\cdot F^i\,_{j}(V)\Bigl]\nonumber\\
[2mm]
& &
- \tilde{\bgl}_R\stackrel{\leftrightarrow}{\sder}\tilde{\gl}_{R} - \frac{1}{2}\Bigl(\frac{1}{2}\bgl^{(ij)}_R\stackrel{\leftrightarrow}{\sder}\gl_{(ij)R} + \frac{1}{2}\bgl^{(ij)}_L\stackrel{\leftrightarrow}{\sder}\gl_{(ij)L}\Bigl) - 
\bgl_{R}^i\,_{j}\stackrel{\leftrightarrow}{\sder}\gl_{R}^i\,_{j}.
\end{eqnarray}
Apart from the scalar $h$, the masses of the  gauge fields can be read off 
easily form the lagrangian $\cL_{\mathrm{bos}}$ given by eq.~(\ref{bos}); they read:
\begin{eqnarray}
\mathrm{m}^2_W = \frac{4}{f^2} g^2_{10}M_0,\quad \mathrm{m}^2_\tA = 
40 g^2_{10}v^2,\quad 
M_0 = (\frac{1}{2 f^2} - 2|v|^2\Bigl) = \frac{\gm^2_1}{10g_{10}^2} > 0.
\end{eqnarray}
By expanding the potential $V_{\mathrm{full}}$ to second order in $\gr$ and $\tk$ 
with scalar $\gr$ defined by
\begin{eqnarray}
 h =(v + \frac{1}{\sqrt{2}}\gr)e^{\frac{1}{\sqrt{2}v}i\ga},
\labl{comscalar}
\end{eqnarray}
around the absolute minimum (\ref{MINIMUM})
we find  
\begin{eqnarray}
V_\mathrm{full} =  V_\mathrm{uni} + \gD V = \frac{1}{2}\mathrm{m}^2_\gr\,\gr^2 
+ \mathrm{m}^2_\tk\,\tk^2 + \dots,
\end{eqnarray}
with $\mathrm{m}^2_\gr = 40 g^2_{10} v^2$ and $\mathrm{m}^2_\tk = 
\frac{1}{f^2}(\frac{3\gm_1^2}{5} + \gm_2^2)$.

Next we construct the fermionic mass terms. The quadratic part of the lagrangian is
\begin{eqnarray}
\cL_{\mathrm{ferm}} &=& - \tilde{\bgl}_R\stackrel{\leftrightarrow}{\sder}\tilde{\gl}_{R} - \frac{1}{4}\Bigl(\bgl^{(ij)}_R\stackrel{\leftrightarrow}{\sder}\gl_{(ij)R} + \bgl^{(ij)}_L\stackrel{\leftrightarrow}{\sder}\gl_{(ij)L}\Bigl) -
\bgl_{R}^i\,_{j}\stackrel{\leftrightarrow}{\sder}\gl_{R}^i\,_{j}\nonumber\\
[2mm]
& &
-  {G_\gs}_{(ij)}{}^{(kl)}\,\bgps^{(ij)}_L\stackrel{\leftrightarrow}{\sder}\gps_{(kl)L} 
- \bgo^i_L\stackrel{\leftrightarrow}{\sder}\go_{iL} - \bgvf_L\stackrel{\leftrightarrow}{\sder}\gvf_{L}\nonumber\\
[2mm]
& &
+ 2\sqrt{2}g_{10}\,{G_\gs}_{(ij)}{}^{(kl)}\,\Bigl[\frac{1}{f}\bgl^{(ij)}_R\gps_{L(kl)} + 
\textrm{h.c.}\Bigl] + 2\sqrt{2}g_{10}\Bigl[2\sqrt{5}v\tilde{\bgl}_R\gvf_L + \textrm{h.c.}\Bigl].
\labl{Fmassso10}
\end{eqnarray}
As a result, two Dirac fermions are  formed by combining the 
quasi-Goldstone fermions $\gps^{[ij]}_L$ and $\gvf_L$ 
with the right-handed gauginos $\gl^{[ij]}_{R}$ and $\tilde{\gl}_R$ according to:
\begin{eqnarray}
\gPs = \tilde{\gl}_R  + \gvf_L,\qquad\gL^{[ij]} = 
\sqrt{M_0}\gps^{[ij]}_L + \frac{1}{2}\gl^{[ij]}_R.
\labl{diracspin}
\end{eqnarray}
In terms of these fields, the fermionic lagrangian becomes
\begin{eqnarray}
\cL_{\rm{ferm}}
&=& - \bgL(\stackrel{\leftrightarrow}{\sder} - \mathrm{m}_{\gL} )\gL - 
\bgPs(\stackrel{\leftrightarrow}{\sder} - \mathrm{m}_{\gPs} )\gPs - 
\bgl_{R}^i\,_{j}\stackrel{\leftrightarrow}{\sder}\gl_{R}^i\,_{j} - \frac{1}{4}\bgl^{(ij)}_L\stackrel{\leftrightarrow}{\sder}\gl_{(ij)L}\nonumber\\
[2mm]
& &  - \bgo^i_L\stackrel{\leftrightarrow}{\sder}\go_{iL},
\end{eqnarray}
with the masses $\rm{m}_{\gL} = \frac{\sqrt{2}\gm_1}{\sqrt{5}f}$ and $\rm{m}_{\gPs}$ 
= $2g_{10}v\sqrt{10}$.
The $\underline{\bar{5}}$  of the left-handed chiral fermions 
$\go_{iL}$, the $\underline{10}$ of the left-handed gaugino's $\gl^{[ij]}_L$, and  the Majorana fermions $\gl^i_{Rj}$ that 
are the gauginos of the unbroken $SU(5)$ symmetry remain massless.
\begin{table}[t]
\begin{center}
\begin{eqnarray}
\renewcommand{\arraystretch}{1.2}
\arry{|l|}{ \hline 
\arry{ c }{ 
~~~~~~~~~~~~~~~~~~~\textrm{fermions} \\
~~~~~~~~~~~~~~~~~~\begin{tabular}{|l|c|c|c|c|c|c|c|} \hline
mass & $\mathrm{m}_{\gPs}^2$&$\mathrm{m}_{\gL}^2$
&$\mathrm{m}_{\gl^{[ij]}_L}^2$&$\mathrm{m}_{\gl^i_j}^2$&$\mathrm{m}_{
\go}^2$\\
[2mm]
\hline
value & $40g^2_{10}v^2$ &$\frac{2\gm^2_1}{5f^2}$&$0$&
$0$&0\\ 
[2mm]
\hline
\end{tabular}
}
\nonumber\\[6ex] 
\arry{c }{ 
\arry{c}{\arry{l}{ 
}}
\arry{c}{ 
\textrm{vectors} \\ 
\begin{tabular}{|l|c|c|c|c|c|} \hline
mass & $\mathrm{m}_{\tA}^2$& $\mathrm{m}_{W}^2$&
$\mathrm{m}_{V}^2$\\
[2mm]
\hline
value & $40\,g^2_{10}v^2$&$\frac{\gm^2_1}{5f^2}$&$0$\\ 
[2mm]
\hline
\end{tabular}
}
\arry{c}{ 
~~~~\textrm{scalars} \\ 
\begin{tabular}{|l|c|c|c|} \hline
mass &  $\mathrm{m}_{\gr}^2$ & $\mathrm{m}_{\tk}^2$\\
[2mm]
\hline
value &$40\,g^2_{10}v^2$ &$\frac{1}{f^2}(\frac{3\gm_1^2}{5} + \gm_2^2)$\\ 
[2mm]
\hline
\end{tabular}
}
}
\\[6ex] \hline 
}
\nonumber
\labl{gaugeso10u1}
\end{eqnarray}
\end{center}
{\sf \caption{Fully gauged $SO(10)$  mass spectrum in the presence of soft supersymmetry breaking.}
\labl{fullygso10withsof1}
}
\end{table}
Notice here that in the limit $\mu^2_{1,2}\rightarrow 0$ and $g_{10} = g_1$, one gets 
the same massive multiplets in the model with only gauged linear subgroup $SU(5)\times U(1)$ (see table \ref{sbreasu5}). The only difference is, 
that in the case of gauged linear subgroup $SU(5)\times U(1)$ there are 20 
massless Goldstone bosons $(\tilde{\bz},\tilde{z})$, and their superpartners 
$(\gps_L,\bgps_L)$; and no gauge bosons $(\bW, W)$ 
of the 20 broken generators of $SO(10)$. (We have observed a similar thing to 
happen also in $E_6/SO(10)\times U(1)$ model discussed in the following section.) From the  massive  spectrum of the theory as summarized in the table \ref{fullygso10withsof1}
, we obtain the general supertrace formula (\ref{sofsum})
\begin{eqnarray}
{\rm STr}\,\mathrm{m}^2 = \mathrm{m}_{\gr}^2 + 2 \mathrm{m}_{\tk}^2 + 3\mathrm{m}_{\tA}^2 + 
6\mathrm{m}_{W}^2 - 4\mathrm{m}_{\gPs}^2 - 4\mathrm{m}_{\gL}^2 = 
\frac{1}{f^2}\Bigl(\frac{4}{5}\gm_1^2 + 2\gm_2^2\Bigl).
\end{eqnarray}
 
Of course, the present theory cannot be regarded as complete. On the one hand, extra 
fermions 
must be coupled to the lagrangian (\ref{comkh}) to represent the other families of 
quarks and leptons. Therefore the  model must  consist of  (at least) three 
copies of $\undr{10}$ , $\undr{\ovr{5}}$ and 
$\undr{1} $ of $SU(5)$ representations in its spectrum, of which one of the 
$\undr{10}$  
are Goldstone bosons of the coset space.  On the other hand,  since one must require the 
remaining $SU(5)$  symmetry to break down at lower energy to $SU(3)\times SU(2)_L\times U(1)$,  additional interactions are required.  For example,  we can add the $\undr{24}$  representation of $SU(5)$ to break  
$SU(5)$ down to smaller symmetry 
group, which can still accommodate at least unbroken $SU(3)\times U(1)$.  
However, 
the symmetry 
breaking in $SU(5)$-GUT via the $\undr{24}~ (\gF)$ that 
acquires a v.e.v. of the form
\begin{eqnarray}
\langle \gF \rangle = 
\textrm{diag}\left( 
v, v, v, - \frac 32 v, - \frac 32 v
\right),
\end{eqnarray}
is problematic. This is because the Higgs-doublets and 
Higgs-triplets, originating from the $\undr{5}$ and 
$\ovr{\undr{5}}$ representations will naturally have  
almost the same effective mass. Now these masses should be very 
large in order to avoid proton decay but on the other hand small,  
else the standard model Higgses are far too heavy. This inconsistency 
is called the doublet-triplet-splitting problem. A way out of this problem is provided 
by the Dimopoulos-Wilczek mechanism \cite{Dimopoulos:1982} as is 
discussed in ref.\ \cite{Babu:1993we}, and recently by Witten \cite{witten}. Such an 
analysis of including other families of 
quarks and leptons as well as additional interactions to break  
$SU(5)$ down to the standard model gauge group is outside the scope of this 
paper and 
requires further development. 
For the moment we are satisfied  with the observation that it is at least possible to 
cure some of the 
difficulties mentioned above for the present model with the scalar particle content summarized in table \ref{so10u5Content}  in principle. 
\begin{table}[t]
\begin{center}
\begin{eqnarray}
\arry{ c }{ 
\begin{tabular}{|l|c|c|c|c|c|} \hline
\textrm{Dimension}  & U(1) &\textrm{Notation} & 
\textrm{Description of the type of fields}\\
repr. & charges & & \\ 
[2mm]
\hline
10 & $ 1 $ &$z^{ij}$ & \textrm{$SO(10)/[SU(5)\times U(1)]$ coset coordinates}\\ 
[2mm]
$\ovr{5}$ & -3 & $k_i$ &
\textrm{Matter additions to $\undr{10}$}\\
[2mm]
1 & 5 & $h$&
\textrm{to complete the $\undr{16}$}\\
\hline
10 & $ 1 $ &$x^{ij}$ & \\ 
[2mm]
$\ovr{5}$ & -3 & $v_i$ &
\textrm{Second family}\\
[2mm]
1 & 5 & $a$&\\
\hline
10 & $ 1 $ &$y^{ij}$ & \\ 
[2mm]
$\ovr{5}$ & -3 & $n_i$ &
\textrm{Third family}\\
[2mm]
1 & 5 & $h$&\\
\hline
24 & 0 & $s^i\,_j$ & 
\textrm{Higgs for breaking the $SU(5)$}\\
[2mm]
 &  &  &
\textrm{group to the standard model}
\\ 
\hline
$\ovr{5}$ & -2 & $c_i$ &
\textrm{ Higgses for breaking the $G_{SM}$}\\
[2mm]
5 & 2 & $c^i$&\textrm{group to the $SU(3)\times U(1)$}\\
\hline
\end{tabular}
}
\nonumber
\end{eqnarray}
\end{center}
{\sf \caption{The various $SU(5)$ representations used for our construction 
of a phenomenological model build around $SO(10)/[SU(5)\times U(1)]$. The first 
column gives the dimension of the representations, the second 
column their charges, the third column the notation we use
for the scalar components of chiral multiplets. A brief description of what these 
fields are is given in the last column. 
\labl{so10u5Content}
}}
\end{table}
\begin{table}[h!]
\begin{center}
\begin{eqnarray}
\renewcommand{\arraystretch}{1.2}
\arry{|l|}{ \hline 
\arry{ c }{ 
~~~~~~~~~~~~~~~~~~~~~~~\textrm{scalars} \\
~~~~~~~~~~~~~~~~~~~~~~~~~~\begin{tabular}{|l|c|c|c|c|c|} \hline
mass &  $\mathrm{m}_{\gr}^2$ & $\mathrm{m}_{\tk}^2$&$\mathrm{m}_{\tz}^2$
 \\ 
[2mm]
\hline
value & $40\,g^2_{1}v^2$ &$0$&$0$\\ 
[2mm]
\hline
\end{tabular}
}
\\[6ex] 
\arry{c}{ 
\arry{c}{\arry{l}{ 
}}
\arry{c}{ 
\textrm{vectors} \\ 
\begin{tabular}{|l|c|c|c|c|} \hline
mass & $\mathrm{m}_{\tA}^2$& $\mathrm{m}_{V}^2$
 \\ 
[2mm]
\hline
value & $40\,g^2_{1}v^2$ &0\\ 
[2mm]
\hline
\end{tabular}
}
\arry{c}{ 
~~~~\textrm{fermions}\\ 
\begin{tabular}{|l|c|c|c|c|c|} \hline
mass &  $\mathrm{m}_{\gPs}^2$ &$\mathrm{m}^2_{\go_{iL}}$&$\mathrm{m}_{\gps^{(kl)}_L}^2$ &$\mathrm{m}_{\bgl_{R}^i\,_{j}}^2$\\
[2mm]
\hline
value & $40\,g^2_{1}v^2$ &0&0&0\\ 
[2mm]
\hline
\end{tabular}
}
}
\\[6ex] \hline 
}
\nonumber
\end{eqnarray}
\end{center}
{\sf \caption{Supersymmetric gauged $SU(5)\times U(1)$ 
mass spectrum
\labl{sbreasu5}
}}
\end{table}

\subsection{Gauging of the linear subgroup $\bf{SU(5)\times U(1)}$\labl{4.3}}

As an alternative to gauging $SO(10)$, one can gauge  only the linear subgroup 
$SU(5) \times U(1)$ instead. This explicitly breaks the non-linear global $SO(10)$. 
It is then allowed in principle to construct superpotentials which are invariant only 
under the local gauge symmetry.
In addition, when gauging any group containing the $U(1)$ as a factor, 
the introduction of a Fayet-Iliopoulos term is allowed. It turns out, that the 
corresponding 
models are indeed well-behaved for a range of non-zero values of this parameter.  

As the  $SU(5)\times U(1)$ subgroup of $SO(10)$ symmetry is not 
broken in the original $\gs$-model, the Killing vectors corresponding to 
these symmetries are linear in the fields. The gauge covariant derivatives are then the usual 
one:
\begin{eqnarray}
D_\gm h &=& \der_\gm h - 2\sqrt{5}g_1\tilde{A}_\gm h,\quad D_\gm k = \der_\gm k + g_5(V^T_\gm + \sqrt{5}\tilde{A}_\gm )k,
\nonumber\\
[2mm]
D_\gm\varphi_L &=& \der_\gm\varphi_L - 2\sqrt{5}g_1\tilde{A}_\gm\varphi_L,\quad 
D_\gm\go_L= \der_\gm\go_L + g_5(V^T_\gm + \sqrt{5}\tilde{A}_\gm )\go_L.
\labl{gu5}
\end{eqnarray}
To determine the physical realization and the spectrum of the 
theory, we have to minimize the potential (\ref{SU5xU1}). This potential has 
absolute minimum at zero if
\begin{eqnarray}
|z|^2 = | k |^2 = 0, \quad | h |^2 = \frac{1}{4f^2} +  
\frac{1}{10}\gx = v^2,\quad -\frac{5}{2f^2}\leq \xi < 0.
\labl{VEV}
\end{eqnarray}
This solution is supersymmetric and spontaneously breaks $U(1)$, whilst 
$SU(5)$ is manifestly preserved. As a result, the $U(1)$ gauge field 
$\tilde{A}_\gm$ become massive with a mass $\mathrm{m}^2_{\tA_\gm} = \mathrm{m}^2_\gr$, 
the mass of the real scalar $\gr$ defined by (\ref{comscalar}). The 
remaining vectors $V_\gm$ of $SU(5)$ stay massless. 
Of the gauginos, the right-handed components of the $U(1)$ gauge multiplet 
$\tilde{\gl}_R$ combine with the left-handed chiral fermions $\varphi_L$ to 
become massive Dirac fermions with the same  mass as the gauge boson 
$\tilde{A}_\gm$. However, the Majorana fermions $\gl_{R}^i\,_{j}$ that are the gauginos of 
unbroken $SU(5)$ symmetry stay massless. 

To see how this result is obtained in more detail, first notice 
that the mass term of the $U(1)$ vector field is generated through the 
kinetic terms by the v.e.v.\ of $h$, 
and reads 
\begin{eqnarray} 
\mathrm{m}_{\tA}^2 =  40 g^2_1\,v^2
\end{eqnarray} 
Next we construct the kinetic terms and potential for the real scalar 
$\gr$; it reads 
\begin{eqnarray} 
\cL(\gr) = - \frac{1}{2}\Bigl[\der\gr\cdot\der\gr 
 -  40 g^2_1\,v^2\gr^2\Bigl] + ...,
\label{n.3}
\end{eqnarray} 
with $\gr$ defined by equation (\ref{comscalar}). We then find that $\gr$ represents 
a real scalar of mass $\rm{m}_{\gr}^2 = \rm{m}_A^2$, the vector boson mass. 
Finally, the kinetic and mass terms for the fermion fields are given by equation (\ref{Fmassso10}) with $g_{10} = g_1$, but without the gauginos of the 20 broken generator of $SO(10)$ (hence the terms involving $\bgl^{(ij)}_R$ are absent.)
\begin{eqnarray}
\cL_{\rm{ferm}} &=& - \tilde{\bgl}_R\stackrel{\leftrightarrow}{\sder}\tilde{\gl}_{R} -
\bgl_{R}^i\,_{j}\stackrel{\leftrightarrow}{\sder}\gl_{R}^i\,_{j}
-  {G_\gs}_{(ij)}{}^{(kl)}\,\bgps^{(ij)}_L\stackrel{\leftrightarrow}{\sder}\gps_{(kl)L} 
- \bgo^i_L\stackrel{\leftrightarrow}{\sder}\go_{iL} - \bgvf_L\stackrel{\leftrightarrow}{\sder}\gvf_{L}\nonumber\\
[2mm]
& &
+ 2\sqrt{2}g_1\Bigl[2\sqrt{5}v\tilde{\bgl}_R\gvf_L + \textrm{h.c.}\Bigl].
\end{eqnarray}
It follows that  the Dirac spinor 
$\gPs = \tilde{\gl}_R + \gvf_L$ satisfies 
the massive Dirac equation 
\begin{eqnarray} 
(\sder + \mathrm{m}_\gPs) \gPs = 0, 
\label{n.5}
\end{eqnarray} 
with $\mathrm{m}_\gPs^2 =  \mathrm{m}_A^2 = \mathrm{m}_{\gr}^2$. This establishes the presence of a 
massive vector supermultiplet $(A_\gm,\gr, \gPs)$ with mass squared given in table \ref{sbreasu5}.

We end this section by remarking that one can also consider gauging  either 
the $U(1)$ $(g_5 = 0 )$ or $SU(5)$  $(g_1 = 0 )$ symmetry. In the first 
case when gauging only the $U(1)$ symmetry, the minimum potential is at the 
same point as in the $SU(5)\times U(1)$ gauging. Therefore the above 
discussion applies here and one gets the same spectrum with equal masses 
for the $U(1)$ gauge multiplet. On the other hand, if  only $SU(5)$ is gauged, the potential reaches its
 minimum at $z = k= 0$. Then no supersymmetry breaking or internal symmetry 
breaking occurs and all particles in the theory are massless.

\section{Analysis of particle spectrum of {\boldmath $\bf{ E_6/SO(10)\times U(1)}$} model
\label{5}}

We turn our attention in this section to another well known model with a 
phenomenologically interesting particle spectrum,
defined by the homogeneous coset space $E_6/SO(10) \times U(1)$ 
\cite{ysj1,YacSaoJv}.  The target manifold $E_6/SO(10) \times U(1)$ is 
parametrized by 16 complex fields $z^{\ga}$ in a chiral superfield $\gF_\ga = 
(z_\ga,\gps_{L\ga}, H_\ga)$ ($\ga = 1, ..., 16 $), 
transforming as a Weyl spinor under $SO(10)$. Their chiral fermion 
superpartners have the quantum numbers of one full generation of 
quarks and leptons, including a right-handed neutrino. To cancel the $U(1)$-anomaly 
the model is extended to a complete $\underline{27}$ of $E_6$. According to the 
branching rule:
$\underline{27}\, \rightarrow\, \underline{16}(1) + 
 \underline{10}(-2) + \underline{1}(4)$, where the numbers in parentheses 
denote the relative $U(1)$ 
weights. With this choice of matter content, the cancellation of 
chiral anomalies of the full $E_6$ isometry group is achieved \cite{SJ1} by 
introducing a 
superfield $\gPs_m = (N_m, \gc_{Lm})$ ($m = 1,\dots, 10$) which is equivalent to a
$\underline{10}$ of $SO(10)$ with $U(1)$ 
charge -2; and finally a singlet $\gL =(h, \gc_L)$ of $SO(10)$, with $U(1)$ 
charge +4.

The anomaly-free supersymmetric $\gs$--model 
on $\ESO$, is defined by three chiral superfields $(\gF_\ga, \Psi_m, \gL)$ with \Kh\ 
potential given by
\begin{eqnarray}
\cK(\gF,\bgF; \gPs,\bgPs; \gL,\bgL) = K_\gs  + e^{-6f^2K_\gs}|h|^2 + 
g_{mn}\bN_m N_ne^{6f^2K_\gs},
\labl{7.1}
\end{eqnarray}
with $K_\gs = \bz .[Q^{-1}\ln (1 + Q)].z$, the $\gs$-model \Kh\  potential. We have introduced a constant $f$ with the dimension $m^{-1}$, determining 
the scale 
of symmetry breaking $E_6\rightarrow SO(10)\times U(1)$. The positive definite matrix $Q$ 
is defined as
\begin{eqnarray}
Q_\ga\,^\gb = \frac{f^2}{4}M_{\ga\gg}^{\gb\gd}\bz^{\gg}z_\gd,\quad M_{\ga\gg}^{\gb\gd} = 3\gd_\ga^{+\gb}\gd_\gg^{+\gd} - 
\frac{1}{2}\gG_{mn\ga}^{+\,\,\,\,\,\,\gb}\gG_{mn\gg}^{+\,\,\,\,\,\,\gd}.
\end{eqnarray}
Here $\gG_{mn}^+  = \gG_{mn}\gd^+$ are the generators of the $SO(10)$ 
on positive chirality spinors of $SO(10)$ \cite{ysj1}, and $\gd^+$ is the 10-D positive 
 chirality projection operator. Furthermore $g_{mn}$ is the induced metric for the 10-vector 
representation defined by 
\begin{eqnarray}
g_{mn} = \frac{1}{16}\tr\Bigl(g_T(\gS_m C)^\dag g_T(\gS_n C)\Bigl)
\quad\textrm{and}\quad g_T = (\Id_{16} + Q)^{-2}.
\end{eqnarray}
The lagrangian constructed from the \Kh\ potential (\ref{7.1}) is invariant under 
a set of holomorphic Killing vectors generating a non-linear representation 
of $E_6$:
\begin{eqnarray}
\gd z_\ga &=& \frac{i}{2}\gth\sqrt{3}z_\ga 
- \frac{1}{4}\go_{mn}(\gG^+_{mn}\cdot z)_\ga + \frac{1}{2}\Bigl[
\frac{i}{f}\ge_\gb\gd^\gb_\ga 
- \frac{if}{4}\bge^\gb M^{\gg\gd}_{\ga\gb}z_\gg z_\gd\Bigl],
\nonumber\\
[2mm]
\gd h &=& 2 i \Bigl(\sqrt{3}\gth - 3 f \bge\cdot z\Bigl)h,\nonumber\\
[2mm]
\gd N_n &=& - i\sqrt{3}\gth N_n - \go_{nm}N_m - i f\bge\cdot (\gG^+_{mn} - 
3\gd^+_{mn})\cdot z N_m
\labl{killE6}
\end{eqnarray} 
where $\gd^+_{mn} = \gd_{mn}\gd^+$, and  $\gth$, $\go_{mn}$ and $\ge_\ga, \bge^\ga$  are the 
infinitesimal parameters of the $U(1)$, $SO(10)$ and broken 
$E_6$ generators respectively. The corresponding Killing potentials are 
\begin{eqnarray}
\cM_i = 
M_i\,E - \frac{1}{8}e^{6 K_\gs}M_{i,\ga}^{~~\gb}\,g_{T\gg}^{~~\gd}(C\bgS_m)^{\ga\gg}
(\gS_n C)_{\gb\gd}\bN_m N_n,
\end{eqnarray}
with $E$ and the $\gs$-model Killing potentials $M_i = (M_\gth, M^{(mn)}, \bM^\gb, M_\gb )$ given by 
\begin{eqnarray}
M_\gth &=& \frac{1}{f^2\sqrt{3}} - \frac{1}{2}\sqrt{3}\bz^\ga K_{\gs,\ga}
,\quad M^{mn} = - \frac{i}{2}\bz^\ga\gG^{+}_{mn\ga}\,^\gb K_{\gs,\gg}\nonumber\\
[2mm]
\bM^\gb  &=& -\frac{1}{f}K_{\gs,}\,^\gb,\quad M_\gb = -\frac{1}{f}K_{\gs,\gb},\quad
E = 1 - 6e^{-6 K_\gs}|h|^2 + 6e^{6 K_\gs}g_{mn}\bN_m N_n.
\end{eqnarray}
\nit 
Observe the presence of the constant term in the $U(1)$ Killing 
potential $M^{\gth}$ which is required to close the Lie algebra on 
the Killing potentials. 

\subsection{The gauged model\label{c7s2}} 

Apart from the pure supersymmetric $\gs$--model determined by this 
\Kh\ potential (\ref{7.1}), we consider models in which (part of) the
isometries (\ref{killE6}) are gauged. As the $E_6$ is broken, the Higgs 
mechanism operates as follows: the Goldstone 
bosons $(\bz^\uga, z^\ga)$ are absorbed in 
the longitudinal component of the charged vector bosons, and if the full $E_6$ is 
gauged, we may choose the unitary gauge $\bz^\uga = z_\ga = 0$. To analyze the model 
in this gauge, we introduce the covariant derivatives for
the dynamical fields. The expressions for gauge-covariant derivatives of the 
complex scalar and fermions fields read
\begin{eqnarray}
D_\gm z_\ga &=& \der_\gm z_\ga - g\Bigl(\frac{i}{2}\sqrt{3}z_\ga A_\gm + 
\frac{1}{4}(\gG^+_{mn}z)_\ga A_{\gm(mn)} + \frac{1}{2}(
\frac{i}{f}A_{\ga\gm}
-\frac{if}{4}\bA_\gm^\gb\,M^{\gg\gd}_{\ga\gb}\,z_\gg z_\gd)\Bigl),
\nonumber\\
[2mm]
D_\gm h &=& \der_\gm h - 2 i g\Bigl(\sqrt{3}A_\gm - 
3 f \bA^\ga_\gm z_\ga\Bigl)h,\nonumber\\
[2mm]
D_\gm N_n &=& \der_\gm N_n +  i \sqrt{3} g A_\gm N_n + g A_{\gm(mn)} N_m + i f g \bA_\gm\cdot (\gG^+_{mn} - 
3\gd^+_{mn})\cdot z N_m\nonumber\\
[2mm]
D_\gm\gps_{L\ga} &=& \der_\gm\gps_{L\ga} - g\Bigl(\frac{i}{2}\sqrt{3}
A_\gm\gps_{L\ga} + 
\frac{1}{4} A_{\gm(mn)}\,\gG^+_{mn}\gps_{L\ga} - \frac{if}{4}\bA_\gm^\gb\,M^{\gg\gd}_{\ga\gb}\,z_\gg\gps_{L\gd}\Bigl),\nonumber\\
[2mm]
D_\gm\gc_L &=& \der_\gm\gc_L - 2ig\Bigl(\sqrt{3}A_\gm\gc_L - 3f g f \bA^\ga_\gm(\gps_{L\ga}h + 
\gc_L z_\ga)\Bigl),\\
[2mm]
D_\gm\gc_{Ln} &=& \der_\gm\gc_{Ln} + g \Bigl(2 i\sqrt{3}A_\gm\gc_{Ln} + 
A_{\gm(mn)}\gc_{Lm}
+
i f \bA_\gm\cdot (\gG^+_{mn} - 
3\gd^+_{mn})\cdot (\gps_L N_m + \gc_{Lm} z) \Bigl)\nonumber
\labl{E6gcovd}
\end{eqnarray}
Here we have introduced the notation $(A_{\gm\ga},\bA^\ga_\gm)$ for the 32 
charged gauge fields 
corresponding to the broken $E_6$ transformations; $A_{\gm(mn)}$ and 
$A_\gm$ are the gauge fields for the remaining $SO(10)$ and $U(1)$ 
transformations respectively.

We have now to add the kinetic terms for the vector multiplets. They are of 
the canonical form  
\begin{eqnarray}
\cL_{\mathrm{gauge}} &=& - \frac{1}{2}\Bigl(\bgl^\ga_R\stackrel{\leftrightarrow}{\sDer}\gl_{R\ga} + \bgl^\ga_L\stackrel{\leftrightarrow}{\sDer}\gl_{L\ga}\Bigl)
- \frac{1}{2}\bgl^{(mn)}_R\stackrel{\leftrightarrow}{\sDer}\gl^{(mn)}_{R} - 
\bgl_R\stackrel{\leftrightarrow}{\sDer}\gl_{R}\nonumber\\
[2mm]
& &
- \frac{1}{4}\Bigl(F^2_{\gm\gn} + 
\frac{1}{2}F^{(mn) 2}_{\gm\gn} + \bF_{\gm\gn}{}^\ga 
F_{\gm\gn\ga}\Bigl)  + \frac{1}{2}\Bigl(\bD^\ga\,D_\ga +\frac{1}{2}D^{(mn) 2} + 
D^2\Bigl)
\labl{kinvec}
\end{eqnarray}
Here we have included a factor $\frac{1}{2}$ to correct for double counting 
due to anti-symmetry of the indices $mn$.

Next we couple the gaugino fields to the quasi-Goldstone $\gps^\ga_L$ and 
matter fermions $(\gc^m_L, \gc_L)$ through the Yukawa coupling
\begin{eqnarray}
\cL_{\mathrm{Yuk}} &=&  2\sqrt{2}g\,G_{z_{\ga}\bz^{\gb}}\Bigl[\Bigl(-\frac{i}{2}\sqrt{3}\bz^\ga\gl_R - 
\frac{1}{4}(\bz\cdot\gG^+_{mn})_\ga\bgl_{R(mn)} + \frac{i}{8}f
\bz^\ga\bz^\gd\,M_{\gd\gg}^{\gb\ga}\,\gl_{R\gb} \nonumber\\
[2mm]
& &
- \frac{i}{2 f}\bgl^\ga_R\Bigl)\gps_{L\gb}
\Bigl]
+ 2 \sqrt{2}g\,G_{N_{m}\bN^{n}}\Bigl[\Bigl(i\sqrt{3}\bN_m\bgl_R - \bN_l
\bgl_{R(ml)} + i f \bN_m\bz\cdot
(\gG^+_{mn} \nonumber\\
[2mm]
& &
- 3\gd^+_{mn})\cdot\gl_R\Bigl)\gc_{Ln}\Bigl]
+ 2\sqrt{2}g\,G_{h\bh}\Bigl[- 2 i\bh\Bigl(\sqrt{3}\bgl_R - 3 f \bz\cdot\gl_R\Bigl)\gc_L\Bigl]
\nonumber\\
[2mm]
& &
+ 2\sqrt{2}g\,G_{z_{\ga}\bh}\Bigl[- 2 i\bh\Bigl(\sqrt{3}\bgl_R - 3 f \bz\cdot\gl_R\Bigl)
\gps_{L\ga} + 
\bgc_L\Bigl(\frac{i}{2}\sqrt{3}z_\ga\gl_R \nonumber\\
[2mm]
& & - 
\frac{1}{4}\gl_{R(mn)}(\gG^+_{mn}\cdot z)_\ga\Bigl) 
- \frac{1}{2}\bgc_L\Bigl(i\frac{f}{4}
\bgl^\gb_R\,M_{\ga\gb}^{\gg\gd}\,z_\gg z_\gd - \frac{i}{f}\gl_{R\ga}\Bigl)\Bigl]
\nonumber\\
[2mm]
& &
+ 2\sqrt{2}g\,G_{z_{\ga}\bN{^m}}\Bigl[\Bigl(i\sqrt{3}\bN_m\gl_R + 
\bN_l\gl_{R(ml)} + i f \bN^n\bz\cdot
(\gG^+_{mn} - 3\gd^+_{mn})\cdot\gl_R\Bigl)\gps_{L\ga} \nonumber\\
[2mm]
& & + \bgc_{Lm}\Bigl(\frac{i}{2}\sqrt{3}z_\ga\gl_R - 
\frac{1}{4}\gl_{R(mn)}(\gG^+_{mn}\cdot z)_\ga\Bigl)\Bigl] 
+ \textrm{h.c.}.
\labl{lyuk}
\end{eqnarray}
Here $(G_{z_{\ga}\bz^{\gb}} , G_{N_{m}\bN^{n}},\dots)$ are the second mixed 
derivatives of the \Kh\  metric $G_{I\uI} = \cK_{,I\uI}$,  where $I = (z_\ga, N_n, h)$ 
and $\uI = (\bz^\ga, \bN^n, \bh)$.

Finally, elimination of the auxiliary fields $(D^\ga, D^{(mn)}, D)$ from (\ref{kinvec}) 
leads to the scalar potential 
\begin{eqnarray}
V_D = \frac{g^2}{2}\sum_i[\cM_i]^2 = \frac{g^2}{2}\Bigl(\cM^2_\gth + 
\frac{1}{2}\cM^2_{mn} + \bcM^\gb\cM_\gb\Bigl).
\labl{E6potential}
\end{eqnarray}

\subsubsection{Gauging of the full $\bf{E_6}$ symmetry\label{c7s3}}

In this section, we discuss in some detail the gauging of the full non-linear $E_6$.
In this case as already stated, we can choose to study the model in the  unitary 
gauge in which all the Goldstone bosons vanish: $z^\ga = \bz_\ga = 0$. This implies 
that the broken Killing potentials $\bcM^\gb$ and $\cM_\gb$ vanish automatically, leaving us with $SO(10)$ and $U(1)$ Killing potentials $\cM_\gth$ and $\cM_{mn}$:
\begin{eqnarray}
\cM_\gth &=& \frac{1}{f^2\sqrt{3}} - 2\sqrt{3}|h|^2 + 
\sqrt{3}\sum_m|N_m|^2,\quad \cM_{mn} = -i \Bigl(\bN_m N_n - 
\bN_n N_m\Bigl).
\end{eqnarray}
Then the full potential becomes
\begin{eqnarray}
V_{\mathrm{unitary}} &=& 
\frac{g^2}{2}\Bigl(\frac{1}{f^2\sqrt{3}} - 2\sqrt{3}|h|^2 + 
\sqrt{3}\sum_m|N_m|^2\Bigl)^2 + \frac{g^2}{2}\sum_{m,n}|\bN_m N_n - 
\bN_n N_m|^2.
\labl{fullgE6}
\end{eqnarray}
Observe here that in the unitary 
gauge, the potential contains only the terms that one also 
gets in gauging $SO(10)\times U(1)$. Minimization of the potential leads to the following set of supersymmetric 
minima characterized by the equation
\begin{eqnarray}
|\bN_m N_n - \bN_n N_m|^2 = 0,\quad |h|^2 = \frac{1}{6f^2} + 
\frac{1}{2}\sum_m|N_m|^2.
\labl{singularmin}
\end{eqnarray}
The value of the potential vanishes: $\langle V \rangle = 0$, hence it is the absolute minimum of the potential. From (\ref{singularmin}), it follows that $|h|\neq 0$ and the $U(1)$ gauge symmetry is always broken; a solution 
with $|N_m| = 0$ is possible, preserving $SO(10)$.  However, solutions 
with $|N_m|\neq 0$ breaking $SO(10)$ are allowed, and expected in the 
next stage of the symmetry breaking. For example, $SO(10)$ broken solution can be 
chosen as 
\begin{eqnarray}
f\bN_m = \left(
\begin{array}{cccccccccc}
0&0&0&0&0&0&0&0&0&v_{10}
\labl{n10}
\end{array}
\right),\quad |h|^2 = |v_h|^2 = \frac{1}{6f^2} + \frac{v^2_{10}}{2f^2}.
\end{eqnarray}
Since the complex scalar $N_m$ gets a vacuum expectation value; this breaks 
the internal linear $SO(10)$ symmetry, leaving only $SO(9)$. 
This shows that for gauged $E_6$, 
supersymmetry is always preserved, and therefore, one expects the spectrum of 
physical states fall into supersymmetric multiplets with vanishing mass 
supertrace. Indeed the general mass sum rule (\ref{sumrule1}) leads to
\begin{eqnarray}
{\rm STr}\,\rm{m}^2 &=&  2 g^2\,G^{\uI I}\,\cM_i\,\cM_{i,\uI I} = 0.
\labl{sumruleE6}
\end{eqnarray} 
As we have gauge the full $E_6$ the standard linear Fayet--Iliopoulos 
term is of course absent.

\subsubsection{Softly broken supersymmetry\label{c7soft}}

In this subsection we discuss the particle spectrum of the theory at the 
minimum with $SO(10)$ invariant solution: 
\begin{eqnarray}
|N_m|^2 = 0,\quad |h|^2 = \frac{1}{6f^2}.
\labl{singvacuum}
\end{eqnarray}
This shows that the internal symmetry $SO(10)\times U(1)$ is broken to 
$SO(10)$. However, this solution is not acceptable by itself, as it leads to the 
to the vanishing of the metric of the $\gs$--model fields $G_\ga\,^\gb = 0$ (and hence
the masses of the 32 $E_6$ gauge fields
$A^\ga_\gm$ vanish) 
To see that in more detail, we first recall that the \Kh\ metric derived from the \Kh\ 
potential $\cK$ (\ref{7.1}) in the unitary gauge reduces to the form:
\begin{eqnarray}
G_{I\uI} = \cK_{I\uI} =
\left(
\begin{array}{ccc}
\gd_{\ga}^{~\gb}\Bigl(\frac{1}{f^2} - 6 |h|^2 + 18 |N_m|^2 \Bigl) - 4
\bN_m N_n(\gG_{mn}^+)_\ga\,^\gb&0&0\\
0&\gd_{mn}&0\\
0&0&1
\end{array}
\right).
\labl{completmetric}
\end{eqnarray}
It is not difficult to see that at the minimum (\ref{singvacuum}) the \Kh\ metric of 
the $\gs$-model fields in the upper-left coner of (\ref{completmetric}) vanishes; 
and the four-fermion term $R_{z^\ga\bz^\ugb h\bh}\,\bgps^\ga_R\,\gps^\ugb_L\,\bgc_L\,\gc_R$
diverge, just like in the $SO(10)/U(5)$--spinor model.
Clearly, in this domain the model no longer 
correctly describes the physics of the situation (i.e., the correct vacuum 
and the corresponding spectrum of small fluctuations). Therefore we add soft breaking 
terms to shift the minimum a way from the singular point, as we 
discussed in section \ref{4}. These terms involve mass terms 
of the form (\ref{softbrea}) for scalar fields 
$(N_m, h)$. 
We include an $E_6$--invariant soft supersymmetry breaking scalar mass term for the singlet $h$ and the vector $N_m$:
\begin{eqnarray}
V_{\mathrm{soft}} = \gm^2_1\,e^{-6K_\gs}|h|^2 + \gm^2_2\,g_{mn}\,\bN_m N_n\,e^{6f^2K_\gs},
\labl{softbreatems}
\end{eqnarray} 
The full scalar potential with soft breaking term in the unitary gauge is then:
\begin{eqnarray}
V = V_{\mathrm{unitary}} + \gm^2_1 |h|^2 + \gm^2_2 |N_m|^2.
\labl{vfuul}
\end{eqnarray} 
As the complex scalar transforms only under $U(1)$, we choose the unitary gauge 
for the $U(1)$ symmetry, which allow us to write 
\begin{eqnarray}
h = \Bigl(v + \frac{1}{\sqrt{2}}\gr\Bigl)\,e^{\frac{1}{\sqrt{2v}}i\gk},
\labl{u1unitary}
\end{eqnarray} 
where $\gk$ is the longitudinal component of the massive gauge field 
$A_\mu$. We now determine the mass spectrum of the theory. 
Expanding the potential (\ref{vfuul}) to second 
order in the fluctuations $\gr$ and $\tilde{N}_m$ around the minimum
\begin{eqnarray}
|N_m|^2 = 0,\quad |h|^2 = v^2 = \frac{1}{6f^2} - \frac{\gm^2_1}{12 g^2}\quad\gm^2_1 < 2\frac{g^2}{f^2}
\labl{softvacuum}
\end{eqnarray} 
the bosonic terms in the action then become in the unitary gauge 
\begin{eqnarray}
\cL_{\mathrm{bos}}&=&  - \frac{1}{4}F^2_{\gm\gn}(\tilde{A}) - 
\frac{1}{4}\bF^\ga_{\gm\gn}F_{\ga\gm\gn} - 
\frac{1}{8}F^{(mn)2}_{\gm\gn} - \frac{1}{2}\der\gr\cdot\der\gr 
- \der\tilde{N}_m\cdot\der\tilde{N}_m \nonumber\\
[2mm]
& & - \rm{m}^2_{A^\ga}\bA^\ga\cdot A_{\ga}
- \rm{m}^2_{\tilde{A}}\,\tilde{A}_{\mu}^2  - \rm{m}^2_{A_{mn}}\,A_{\mu(mn)}^2  - \frac{\rm{m}^2_\gr}{2}\, \gr^2 - \rm{m}^2_{\tilde{N}_m}\tilde{N}_m^2 \nonumber\\
[2mm]
& & - V_0 + \dots,
\labl{7.28}
\end{eqnarray}
In this expression, the dots represent interactions of the abelian vector field 
with the scalar $\gr$. In addition, we have absorbed the Goldstone mode $\gk$ in 
the abelian vector by redefining the $U(1)$ gauge field $A_\gm$:
\begin{eqnarray}
A_\gm\rightarrow\tilde{A}_\gm = A_\gm - \frac{1}{2\sqrt{6}gv}\der_\gm\gk.
\end{eqnarray}
The masses of the bosonic fields read: 
\begin{eqnarray}
\mathrm{m}^2_{\tilde{A}} = \mathrm{m}^2_\rho = 24g^2 v^2,\quad\mathrm{m}^2_{A^\ga} 
= \frac{\gm^2_1}{4f^2},\quad\mathrm{m}^2_{\tilde{N}_m} = \frac{1}{f^2}\Bigl(\frac{1}{2}
\gm^2_1 + \gm^2_2\Bigl),\quad\mathrm{m}^2_{A_{nm}} = 0.
\end{eqnarray}
As expected the gauge bosons  $A_{\gm[mn]}$ of the non-broken 
$SO(10)$ symmetry remain massless.

Analyzing the kinetic and mass terms of the fermions 
\begin{eqnarray}
\cL_{\mathrm{ferm}} &= &- G_\ga^{~\gb}\bgps^\ga_{L} \stackrel{\leftrightarrow}{\sder} \gps_{L\gb} - \bgc^n_{L} \stackrel{\leftrightarrow}{\sder} \gc^n_L -  
\bgc_{L} \stackrel{\leftrightarrow}{\sder} \gc_L - \frac{1}{2}\Bigl(\bgl^\ga_R\stackrel{\leftrightarrow}{\sder}\gl_{\ga R} + \bgl^\ga_L\stackrel{\leftrightarrow}{\sder}\gl_{L\ga}\Bigl)\nonumber\\
[2mm]
& &
- \frac{1}{2}\bgl^{(mn)}_R\stackrel{\leftrightarrow}{\sder}\gl^{(mn)}_{R} - 
\bgl_R\stackrel{\leftrightarrow}{\sder}\gl_{R} + \sqrt{2}g\,G_\ga^{~\gb} \frac{i}{f}\Bigl(\bgps^\ga_L\gl_{R\gb} - \bgl^\ga_R\gps_{L\gb}
\Bigl)\nonumber\\
[2mm]
& &
+ 4i\sqrt{6}vg\Bigl(\bgc_L\gl_R -  \bgl_R\gc_L\Bigl)
\labl{lyuk1}.
\end{eqnarray}
one realizes that two massive Dirac fermions can be formed by combining the fermions of the chiral multiplets with two 
gauginos:
\begin{eqnarray}
\gPs_\ga = \frac{1}{\sqrt{2}}\gl_{R\ga} - i\sqrt{2}\,\frac{\gm_1}{2g}\,\gps_{L\ga}, 
\qquad 
\gO = \gl_R - i\gc_L.
\end{eqnarray}
In terms of these fields, the expression (\ref{lyuk1}) becomes
\begin{eqnarray}
\cL_{\mathrm{ferm}} = - \bgPs^\ga\stackrel{\leftrightarrow}{\sder}\gPs_\ga - 
\bgO\stackrel{\leftrightarrow}{\sder}\gO + \sqrt{2}\,\frac{\gm_1}{f}\bgPs^\ga\gPs_\ga 
+ 4\sqrt{6}v\,g\,\bgO\gO.
\end{eqnarray}
The masses of these spinors are:
\begin{eqnarray}
\mathrm{m}^2_{\gPs} = \frac{\gm^2_1}{2f^2},\qquad \mathrm{m}^2_{\gO} = 
24g^2 v^2.
\end{eqnarray}
The $\underline{16}$ of the left-handed gaugino's $\gl_{L\ga}$ and 
quasi-Goldstone fermions $\gc_{Ln}$ remain massless, together with the Majorana fermions $\gl^{mn}$ that are 
gauginos of the unbroken $SO(10)$ symmetry. Therefore, in this model the gaugino 
components $\gl_{L\ga}$ are now to be identified with a family of quarks and leptons, 
rather than the quasi Goldstone fermions themselves. (We have observed a similar thing to happen also in the $SO(10)/U(5)$--spinor model discussed in section \ref{4}.) 
The complete spectrum  of the theory is summarized in table \ref{fullygaugeE6}. 
\begin{table}[t]
\begin{center}
\begin{eqnarray}
\renewcommand{\arraystretch}{1.3}
\arry{|l|}{ \hline 
\arry{ c }{ 
~~~~~~~~~~~~~~~~~~~\textrm{fermions} \\
~~~~~~~~~~~~~~~~~~\begin{tabular}{|l|c|c|c|c|c|c|c|} \hline
mass 
&$\mathrm{m}_{\gPs_{\ga}}^2$
&$\mathrm{m}_{\gO}^2$&$\mathrm{m}_{\gc^n}^2$&
$\mathrm{m}_{\gl_{L\ga}}^2$&$\mathrm{m}_{\gl^{[mn]}}^2$\\ 
[2mm]
\hline
value & $\frac{\gm^2_1}{2f^2}$ &$24g^2v^2$&$0$&
$0$&$0$\\ 
[2mm]
\hline
\end{tabular}
}
\nonumber\\[6ex] 
\arry{c }{ 
\arry{c}{\arry{l}{ 
}}
\arry{c}{ 
\textrm{vectors} \\ 
\begin{tabular}{|l|c|c|c|c|c|} \hline
mass & $\mathrm{m}_{A}^2$& $\mathrm{m}_{A_{[mn]}}^2$&
$\mathrm{m}_{A^\ga}^2$
 \\ 
[2mm]
\hline
value & $24g^2v^2$&$0$&$\frac{\gm^2_1}{4f^2}$\\ 
[2mm]
\hline
\end{tabular}
}
\arry{c}{ 
~~~~\textrm{scalars} \\ 
\begin{tabular}{|l|c|c|c|} \hline
mass &  $\mathrm{m}_{\gr}^2$ & $\mathrm{m}_{\tN}^2$\\
[2mm]
\hline
value &$24g^2v^2$&$\frac{1}{f^2}\Bigl(\frac{1}{2}\gm^2_1 + \gm^2_2\Bigl)$\\ 
[2mm]
\hline
\end{tabular}
}
}
\\[6ex] \hline 
}
\nonumber
\labl{gaugeso10u1}
\end{eqnarray}
\end{center}
{\sf \caption{Fully gauged $E_6$ mass spectrum in the presence of soft supersymmetric breaking.}
\labl{fullygaugeE6}
}
\end{table}

The conclusions that can be drawn from the above analysis may be summarized as follows.
Gauging of the full $E_6$ in the presence of soft supersymmetry 
breaking may lead to a possibly realistic description of the lightest family of quarks 
and leptons. To make it fully realistic three important problems must be solved 
\cite{Mohapatra:1997sp}:
\enums{
\item How to break down the remaining $SO(10)$ symmetry, as required by 
low-energy phenomenology.
\item It should be possible to include (at least) three generations 
of quarks and leptons. 
\item There should be a source of large Majorana masses, so that 
the see-saw mechanism provides the explanation for the small 
neutrino masses.
}
 \begin{table}[h!]
\begin{center}
\begin{eqnarray}
\arry{ c }{ 
\begin{tabular}{|l|c|c|c|c|c|} \hline
\textrm{Dimension}  & U(1) &\textrm{Notation} & 
\textrm{Description of the type of fields}\\
repr. & charges & & \\ 
[2mm]
\hline
16 & $ 1 $ &$z_\ga$ & \textrm{$\ESO$ coset coordinates}\\ 
[2mm]
10 & -2 & $N^m$ &
\textrm{Matter additions to $\undr{16}$}\\
[2mm]
1 & 4 & $h$&
\textrm{to complete the $\undr{27}$}\\
\hline
16 & $q$ & $x^+_\ga$ &
\textrm{Two generations} \\ 
[2mm]
16 & -$q$ & $x^-_\ga$ &
\\ \hline
45 & 0 & $A^{mn}$ & 
\textrm{Higgses for the unification}\\
[2mm]
54 & 0 & $S^{mn}$ &
\textrm{symmetry breaking}\\
[2mm]
210 & 0 & $Q^{mnpq}$ &
\\ \hline 
126 & $r$ & $D^{mnpqr}$ &
\textrm{Higgses for neutrino Majorana masses}\\
[2mm]
$\ovr{126}$ & -$r$ & $E_{mnpqr}$ &
\textrm{and symmetry breaking}
\\ \hline
\end{tabular}
}\nonumber
\end{eqnarray}
\end{center}
{\sf \caption{The various $SO(10)$ representations used for our construction 
of a phenomenological model build around $\ESO$. The first 
column gives the dimension of the representations, the second 
column their charges, the third column the notation we use 
for the scalar components of chiral multiplets. A brief description of what these 
fields are is given in the last column. 
The charges $q,r$ will be fixed by dynamical considerations 
like $SO(10)\times U(1)$ anomaly cancellations and the requirement
that various Yukawa couplings can appear in the superpotential. 
\label{esoContent}
}}
\end{table}
Like in the $SO(10)/U(5)$--spinor model, these problems may be solved by adding 
additional matter multiplets. Let us start with the second problem in the list above. The $\undr{16}$ can accommodate one generation 
of quarks and leptons including the right-handed neutrino. 
Therefore we need at least three copies of this representation 
to account for three families. 
It would be economical (as far as the field content is concerned) 
to use a $\undr{16}$ both as a representation of quarks and 
leptons and as the representation that leads to the symmetry 
breaking 
$S0(10) \lra SU(5)\times U(1)\lra SU(3)\times SU(2)_L\times U(1)$. Therefore a possible solution to the this  
problem is provided by adding the two other fermion 
families as additional matter multiplets $\gF^\pm_\ga = (x^\pm_\ga,\gps^\pm_{\ga L})$ 
carrying opposite $U(1)$ charges so that that the internal symmetry is free of 
anomalies. 

The first  problem above can be solved by introducing the $SO(10)$ breaking Higgs 
multiplets $A^{mn}$, $S^{mn}$ and $Q^{mnpq}$ with $U(1)$ charges taken to be zero. This is  
not strictly necessary but very convenient in the following. 
The fermionic partners of the coset coordinates $z^\ga$ 
form one family of quarks and leptons, the other two family
multiplets have scalar components $x^\pm_\ga$. 
We make the charge convention such that $x_+$ has positive 
charge $q \geq 0$. 
Finally, we have 
two additional Higgses $E_{mnpqr}$ and $D^{mnpqr}$ that 
may also be responsible for symmetry breaking, but in addition 
are also supposed to give rise to Majorana masses for the 
right-handed neutrinos. $D$ has charge $r$ and $E$ is it 
charge conjugate. In addition to all this there should be at least a $\undr{10}$ 
that can produce the supersymmetric standard model Higgses 
after symmetry breaking down to the standard model group 
$SU(3)\times SU(2)_L\times U(1)$.

\subsection{Gauging of $\bf{SO(10)\times U(1)}$ symmetry\label{c7s4}}

The gauging of the $SO(10)\times U(1)$ symmetry instead of the full $E_6$ 
gives analogous, but not quite identical, results. Also in this case one 
finds 
the potential (\ref{fullgE6}), but in general with different values $g_1$ and 
$g_{10}$ for the coupling constants of $SO(10)$ and $U(1)$. Except for 
special values of the parameters, it has a minimum for the $SO(10)$ invariant 
solution, with $z_\ga = 0$; and again the metric becomes singular. One way 
to shift the minimum away from this point is by introducing soft breaking 
terms (\ref{softbreatems}). Another option is to add an extra 
Fayet-Iliopoulos term as the gauge group possesses an explicit $U(1)$ 
factor.  In the first case, the fermionic mass term is given by the last  line of 
(\ref{lyuk1}). As a result there is now one massive Dirac fermion, from the 
combination of $\gc_L$ with the same gaugino of the broken $U(1)$ as before. 
The gauginos $\gl^{mn}$ that are left over remain unpaired, and hence 
massless. Furthermore, the chiral fermions  $\gps^\ga_L$ and $\gc^n_L$  
remain 
massless. The complete spectrum can be read from the table \ref{gaugeso10u1}

In the second case, for special values of the coupling constants 
$g_1$ and $g_{10}$, or 
the Fayet-Iliopoulos parameter $\gx$, one can get different results. 
Since the 
$SO(10)$ and $U(1)$ coupling constants are independent, one may choose to gauge only  
$SO(10)$ ($g_1 =  0$). In that case both supersymmetry and  internal symmetry 
are preserved, and the particle spectrum of a model contains of a massless $SO(10)$ 
gauge boson, just like in the usual supersymmetric $SO(10)$ grand unified models.   
\begin{table}[here]
\begin{center}
\begin{eqnarray}
\renewcommand{\arraystretch}{1.3}
\arry{|l|}{ \hline 
\arry{ c }{ 
~~~~~~~~~~~~~~~~~~~~~\textrm{scalars} \\
~~~~~~~~~~~~~~~~~~~~~\begin{tabular}{|l|c|c|c|c|c|} \hline
mass &  $\mathrm{m}_{\gr}^2$ & $\mathrm{m}_{\tN_m}^2$&$\mathrm{m}_{\tz^\ga}^2$
 \\ 
[2mm]
\hline
value & $24g_1^2 $ &$0$&$0$\\ 
[2mm]
\hline
\end{tabular}
}
\\[6ex] 
\arry{c }{ 
\arry{c}{\arry{l}{ 
}}
\arry{c}{ 
\textrm{vectors} \\ 
\begin{tabular}{|l|c|c|c|c|} \hline
mass & $\mathrm{m}_{A}^2$& $\mathrm{m}_{A_{mn}}^2$
 \\ 
[2mm]
\hline
value & $24 g_1^2 v^2 $&0\\ 
[2mm]
\hline
\end{tabular}
}
\arry{c}{ 
~~~~\textrm{fermions} \\ 
\begin{tabular}{|l|c|c|c|c|} \hline
mass & $\mathrm{m}_{\gL}^2$&$\mathrm{m}_{\gps_{L\ga}}^2$
&$\mathrm{m}_{\gc_{Ln}}^2$
 \\ 
[2mm]
\hline
value &$24 g_1^2 v^2$&0&0\\ 
[2mm]
\hline
\end{tabular}
}
}
\\[6ex] \hline 
}
\nonumber
\labl{gaugeso10u1}
\end{eqnarray}
\end{center}
{\sf \caption{Soft supersymmetry breaking gauged $SO(10)\times U(1)$ mass spectrum}
\labl{gaugeso10u1}
}
\end{table}

\section{Conclusions\label{6}}

\nit
The \Kh\  manifolds $\ESO$ and $SO(10)/SU(5)\times U(1)$ hold some 
special interest in the context of non-linear supersymmetric $\gs$-models, because 
$E_6$, $SO(10)$ and $SU(5)$ are realistic grand unification groups. It 
was shown \cite{SJ1,STJ} that it is possible to 
construct anomaly free models around these coset-spaces that 
are globally consistent. 

In this article, we have discussed in detail the phenomenological analysis of 
supersymmetric $\gs$-models on homogeneous  coset-spaces $\ESO$ 
and $SO(10)/U(5)$. We have analyzed the possible vacuum configurations of 
these models. We have investigated in particular the existence of the zeros 
of the potential, for which the models are anomaly-free, with positive definite 
kinetic energy.  The consequences of these physical requirements have been 
analyzed. We found that there exist supersymmetric minima for both these models when the full isometry groups $E_6$ and $SO(10)$ are gauged. The analysis is straightforward 
as one can employ the unitary gauge to put the Goldstone bosons to zero. In some cases, we find that the \Kh\ metrc is singular: the kinetic energy of the would-be Goldstone modes and their fermionic partners vanishes in the vacuum. We showed by addition of soft supersymmetry-breaking mass parameters, that 
the minimum can be shifted away from the singular point.

The particle spectrum in the presence of soft supersymmetry-breaking mass parameters is computed. The gauge bosons corresponding to the broken $E_6$ as well as the $SO(10)$ become massive, thereby eliminating all the Goldstone 
scalars from the theory. In addition some of the left-handed quasi-Goldstone fermions become massive by combining with right-handed gauginos 
corresponding to the broken ($SO(10)$,  $E_6$) generators. The left-handed of these gauginos components remain massless and  have the same quantum number as the original quasi-Goldstone fermions. Therefore, they can represents a family of quarks and leptons, with additional right-handed neutrino.

Continuing our line of investigation of the particle spectrum of supersymmetric 
$\gs$-models on $\ESO$, and $SO(10)\times U(1)$, we have also studied the 
possibility of gauging (part of)  the linear subgroups, i.e., $SO(10)\times U(1)$ 
and $U(5)$. In each of these models, we found that the  properties of the 
model investigated depend to a certain extent on the value of 
parameters (gauge couplings, Fayet-Iliopoulos term) and the presence of 
extra families and Higgses. We have obtained all supersymmetric minima, of which some are 
physically problematic as the kinetic terms of the Goldstone multiplets either 
vanish or have negative values. 

In spite of all these nice features, there is still a lot of work needed to 
improve and extend the anomaly-free supersymmetric $\gs$-models on 
the  coset spaces $SO(10)/U(5)$ and $\ESO$ discussed here. For example, it would be to interesting to  study their particle spectrum
 in presence of extra families and Higgses.  In tables \ref{so10u5Content} and 
\ref{esoContent} we have summarized the most general scalar field content 
we consider for the phenomenological promising models build around the 
coset spaces $SO(10)/U(5)$ and $\ESO$.\\ 

\nit
{\bf Acknowledgements}\nl

\nit
I thank J.W.\ van Holten and S. Groot Nibbelink for very useful, pleasant discussions 
and for reading the manuscript. Most of this work was performed at NIKHEF as part of 
the research programme Theoretical Subatomic Physics (FP52) of the Foundation for 
Fundamental Research of Matter (FOM) and the Netherlands Organization for Scientific 
Research (NWO).

\end{document}